
\magnification=1200
\baselineskip=14pt     
\nopagenumbers

\voffset=1truecm 
\hsize= 5.25in 
\vsize=7.5in 

\def\nl{\par\noindent}
\def\|{\'\i}

\rightline {ISSN 0029-3865}
\vskip .3truecm

\rightline {CBPF-NF- 075/93}
\vskip 1truecm

\pretolerance=10000    

\centerline {\bf LIGHT-FRONT QUANTIZED FIELD THEORY: ({\sl an introduction})
\footnote{*}{{\it
{\bf Lecture} delivered at {\bf XIV Encontro Nacional de Part\|culas e Campos},
Caxambu, MG, September 1993}. {\sl To be published in the Proceedings.}}}\par
\medskip
\centerline {\bf Spontaneous Symmetry Breaking.
Phase Transition in $\phi^{4}$ Theory.}

\vskip 1.5truecm

\centerline {Prem P. Srivastava\footnote{**}{{\it Postal address: Rua Xavier
Sigaud 150, 22290 Rio de Janeiro, RJ, Brasil.}
\nl {\bf e-mail:} prem@cbpfsu1.cat.cbpf.br}}
\medskip
\nl \centerline {\it Centro Brasileiro de Pesquisas F\|sicas, Rio de Janeiro.}

\vskip 2truecm

\centerline  {\bf Abstract}
\medskip

{\leftskip 30pt \rightskip 30pt {\sl \qquad The
field theory quantized on the {\it light-front} is compared
with the  conventional equal-time quantized theory.
The arguments based on the {\it microcausality} principle
would imply that the light-front field theory may become nonlocal
with respect to the longitudinal coordinate even though the corresponding
equal-time formulation is local. This is found to be the case for the scalar
theory. The conventional {\it instant form} theory is sometimes required
to be constrained
by invoking external physical considerations; the analogous conditions seem
to be already built in the theory on the  {\it light-front}.  In spite of the
different mechanisms  of the spontaneous symmetry breaking in the
two forms of dynamics they result in the same physical content.
The phase transition in {$(\phi^{4})_{2}$}
theory is also  discussed. The symmetric vacuum state for vanishingly small
couplings is found to turn into an unstable symmetric one when the coupling is
increased and may  result in a phase transition of
the {\it second order} in contrast to the first order transition
concluded from the usual variational methods.}\par}

\vfill \eject

\nopagenumbers

\headline={\hskip 2.3in\tenrm{ --\folio --}\hfil CBPF-NF-075/93}
\pageno=1

\topskip=20pt   

\baselineskip= 13pt 

\nl {\bf 1- Introduction:}

\medskip
The possibility of building dynamical theory of a physical system on
the three dimensional hypersurface in space-time formed by a plane wave
front advancing with the velocity of light
was indicated by Dirac [1] in 1949.
The initial conditions on the
dynamical variables are now specified
on the hyperplane ({\it light-front}), say, $\,x^{0}+x^{3}=0,\,$ which
has a light like normal,
 the {\it front form},
in contrast to the usual formulation where
we employ instead the $x^{0}\equiv t=0$
hyperplane,  the {\it instant form}. He also argued that the light-front
formulation should be simpler since seven of the ten generators of the
Poincar\'e group  turn out to be kinematical while in the case of the
{\it instant form}
there are only six of them.
The kinematical generators correspond to the
ones which leave the chosen hyperplane invariant. Latter in 1966
the {\it front form} dynamics was rediscovered by Weinberg [2] in the
infinite momentum frame rules in the
quantized field theory which were clarified by Kogut and Soper [3] in 1970
to be equivalent to the quantization on the light-front. Even earlier [4]
the ${ p\to\infty}$ technique played an important role in the derivation of
the current algebra sum rules and it was observed [5] that it amounted to
using appropriate light-front current commutators.
The {\it front form} coordinates are
also adopted frequently in the string theories [6] in
order to be able to work with the
physical degrees of freedom and to expose clearly the physical contents.

A  remarkable feature of the theory quantized on the light-front is
the apparent simplicity of the vacuum state.
In many theories the interacting theory vacuum coincides with that
of the perturbation (free) theory one. In fact,
the four momentum components are now $(k^{-},k^{+},k^{\perp})$ where
$k^{\pm}=(k^{0}{\pm}k^{3})/{\sqrt 2}$ and $\,{k^{\perp} }\equiv \bar k=(
k^{1},k^{2})\,$ indicate the two
components transverse to $x^{3}$-direction. For a massive free
particle on its mass shell and $k^{0}>0$ we find
$k^{\pm}$ nonvanishing and positive. On the other hand
in the {\it instant form}, the momentum
eigenstates of a particle is specified by the components
$(k^{1},k^{2},k^{3})$ which may take positive or negative values. We may
construct here eigenstates of zero momentum with arbitrary number of
particles (and antiparticles) which may mix with the vacuum state, without
any particle, to form the ground state. In contrast in the light-front
framework we require $k^{+}\to 0$ for each of the particle entering the
ground state with vanishing total momentum. Such configurations constitute
a point with zero measure in the phase space and may not be of relevance
in many cases. It should, however, be remarked that
when dealing with momentum space integrals,
say, the loop integrals, in some cases a significant contribution may
arise precisely from such a (corresponding) configuration in the integrand;
the reason being that we have to deal with products of several distributions.

The recent revival of interest [7,8] in the light-front
theory has been motivated
by the difficulties faced in the nonperturbative QCD- the gauge theory
of quarks and gluons- in
the usual {\it instant formulation}. The technique of the
regularization on the lattice has been
quite successful for some problems but it cannot handle, for example, the
light ( or chiral fermions) and  has not  been able yet
to demonstrate confinenment of
the quarks.
We have also the open problem of reconciling
the standard constituent quark model and QCD to
describe the hadrons [7,8]. In the former
we employ few valence quarks while in the latter the QCD vacuum state itself
contains an infinite sea  of constituent quarks and gluons ( partons)
with the density
of low momentum constituents getting very large in view of infrared
slavery. Another problem is that of relativistic bound-state computation
in the presence of the complicated vacuum in the {\it instant form}.
Recent studies [7,8] show that the application of
Light-front Tamm-Dancoff method [7] may be feasible
here.  The front-form dynamics may serve as a
complementary tool where we have a simple vacuum while the
complexity of the problem is now transferred to the
light-front Hamiltonian. In the case of the scalar field theory, for example,
discussed below the corresponding light-front  Hamiltonian is found [9,10]
to be nonlocal due to the presence of
{\it constraint equations} in the Hamiltonian formulation. A different
description of the spontaneous symmetry breaking is obtained which is,
however,  equivalent ([9], Sec. 2) in the physical contents
to the usual description in the
{\it instant form}. In the latter case we customarily do add to the theory some
physical requirements from outside while such conditions seem to be already
incorporated in the light-front context through the self-consistency
requirements and the constraint equations (Sec. 2). We give arguments
that the {\it nonlocality}  mentioned above is not unexpected
and it {\it does not enter into conflict with the microcausality principle}.

A general feature of the {\it front form} theory is that it describes
a constrained dynamical system and the construction of Hamiltonian
formulation is not straightforward. The Dirac procedure [11]
or its variants must be used to handle such a system.
Selfconsistent classical Hamiltonian formulation [1] is very convenient
to quantize the theory via the correspondence of
the Dirac (Poisson) brackets with
the commutators of the corresponding operators and also allows us to
unify the principle of (special) relativity in the dynamical theory
again by making use of these brackets.

We introduce the following notation.  For the coordinate  $ x^{\mu}$ and
for all other vector or tensor quantities we define the $\pm$ components
$x^{\pm}=(x^{0}{\pm} x^{3})/{\sqrt 2}=x_{\mp}$. We adopt  $x^{+}$
to indicate the {\it light-front time coordinate} and $ x^{-}$ the
{\it spatial longitudinal coordinate}.
The {\it spatial transverse} components will be usually
denoted by ${\bar x}\equiv x^{\perp}=(x^{1}=-x_{1},\,x^{2}=-x_{2})$.
The metric tensor for the
indices $\mu=(+,-,1,2)$ is given by $g^{++}=g^{--}=g^{12}=g^{21}=0; \,
g^{+-}=g^{-+}=-g^{11}=-g^{22}=1\,$ and it is verified that $g_{\mu\nu}A^{\mu}
B^{\nu}=A^{\mu}B_{\mu}=A^{-}B^{+}+A^{+}B^{-}-A^{1}B^{1}-A^{2}B^{2}$ is
the correct Lorentz invariant expression, for example,
$x^{\mu}x_{\mu}\equiv x^{2}=2x^{+}x^{-}-{\bar x}^{2}$.
Under the pure Lorentz transformations
in $(0,3)$ plane we note that the components $A^{\pm}$ undergo scale
transformations such that both $A^{+}B^{-}$ and $A^{-}B^{+}$ are left
invariant. Their sum corresponds to the usual invariant $A^{0}B^{0}-
A^{3}B^{3}$,  while the difference to the invariant
$A^{0}B^{3}-A^{3}B^{0}$ which has a symplectic structure.
It is easily verified that the transformation from the usual coordinates
$(x^{0},x^{3},{\bar x})$ to the coordinates
$(x^{+},x^{-},{\bar x})$ is {\it not} a Lorentz transformation.

It is well known that two distinct points lying on the hyperplane $x^{0}=
const.$ are separated by space like distance, e.g., $(x-y)^2=-(\vec x
-\vec y)^{2}< 0$, and the
separation becomes light like when the two points become coincident. The
points on the hyperplane $x^{+}=const.$ also have space like separation for
$x^{\perp}\ne y^{\perp} $. It becomes light like when
$x^{\perp}=y^{\perp}$, however, with
the {\it difference} that the
points now need not become coincident, since  $(x^{-}- y^{-})$ is not
required to be vanishing. This observation
when combined with the {\it microscopic causality}
postulate: `{\sl the commutators of two physical observables pertaining to
space-time points which are separated by a space like distance be
vanishing}', leads to the result that
the {\it front form} (quantized ) dynamics may become non-local with respect to
the longitudinal (space) coordinate $x^{-}$.
Consider, for example, the commutator
$[A(x^{0},{\vec x}),B(0,{\vec 0})]$ of two scalar observables
$A(x)$ and $B(x)$ where $\vec x$ indicates the usual 3-vector
(in equal-time formulation). It is function of the invariant
$x^{2}$ due to Lorentz invariance and vanishes
for  $x^{2}<0$ if microcausality condition is assumed.
Employing the light-front coordinates and evaluating the commutator
on the light-front we find that
$[A(x^{+},x^{-},{\bar x}),B(0,0,\bar 0)]_{x^{+}=0}$ should vanish
for $\bar x \ne 0$, since $x^{2}=-{\bar x}^2<0$ when  $x^{+}=0$. This
commutator hence is non-vanishing  only for ${\bar x}=0$ when
also $x^{2}$ becomes light-like. We thus expect  that its value
contains a $\delta^{2}(\bar x)$ and its derivatives  which would
imply locality in $\bar x$. No constraint, however, is obtained
on the $x^{-}$ dependence which is arbitrary.
In the {\it instant form} case similar arguments applied
to the equal-time commutator  $[A(x^{0},{\vec x}),B(0,\vec 0)]_{x^{0}=0}$
lead to the possible presence of $\delta^{3}(\vec x)$ and its derivatives
implying locality in all the three space coordinates.
We remark that in view of the
microcausality the knowledge of the
equal-$x^{+}$ or equal-$x^{0}$ commutator is equivalent to finding
its value on the light-cone $x^{2}=0$,
while approaching it from the space like regions.

In order to obtain some information on the nature of the light-front
commutator, say,  of the scalar field,
we may consider the corresponding {\it Lehmann spectral representation}
[12] for the vacuum expectation value of the commutator given by

$${\langle 0\vert [\phi(x),\phi(0)]\vert 0\rangle}= \int_{0}^{\infty}\,
d\sigma^2\,\rho(\sigma^2)\,\triangle(x;\sigma^2).\eqno(1.1)$$

\nl Here $\phi$ is is an Heisenberg operator,
$\rho(\sigma^2)$ is Lorentz invariant
positive-definite spectral function, and
$\triangle(x;\sigma^2)$ is the free field commutator function

$$\triangle(x;\sigma^2)={1\over {(2\pi)^3}} \int_{-\infty}^{\infty}\,
 d^{4}k\epsilon(k^{0})
\delta(k^2-\sigma^2)\,e^{-ik.x}\,\eqno(1.2).$$

\nl where the distribution $\epsilon(y)=-\epsilon(-y)=\theta(y)-\theta(-y)
=1 $ for $y>0$.
In case the theory is derived from a local Lagrangian we may
also establish, by making use of the canonical equal-time
commutation relations, the result

$$\int_{0}^{\infty}d\sigma^2\,\rho(\sigma^2)=1\,\eqno(1.3).$$

\nl Let us compute the free field commutator (2) on the light-front
$x^{+}=0$. We note
that $d^{4}k=d^{2}{\bar k}dk^{+}dk^{-},\,k^2=2k^{+}k^{-}-{\bar k}^2,
\, k.x=k^{+}x^{-}+k^{-}x^{+}-{\bar k}.{\bar x},\, and
(2\vert k^{+}\vert)\delta(k^2-\sigma^2)=
\delta(k^{-}-[{\bar k}^2+\sigma^2]/(2k^{+}))$. In view of the mass shell
condition implied by the delta function
it is easily shown that $(k^{-}/k^{+})>0$ and from the definition
$k^{0}=(k^{+}+k^{-})/{\sqrt 2}$ it follows that inside the integral
$\epsilon(k^{0})=\epsilon(k^{+})$. On setting  $x^{+}=0$ and
integrating  over $k^{-}$ (to remove the delta function) and
$\bar k$ we obtain

$$\triangle(x^{+},x^{-},{\bar x};\sigma^2)\vert_{x^{+}=0}
=-{i\over 4}\delta^{2}(\bar x)\epsilon(x^{-}),\,\eqno(1.4)$$

\nl which does not depend on $\sigma^{2}$. On using (3) we obtain

$$ [\phi(x^{+},x^{-},{\bar x}),\phi(0)]\vert_{x^{+}=0}=
-{i\over 4}\delta^{2}(\bar x)\epsilon(x^{-}),\,\eqno(1.5)$$

\nl as far as the vacuum expectation value is concerned. We will give below
an independent derivation of this light-front commutator
by quantizing the scalar field theory directly
in the {\it front form} by following the Dirac procedure. From (1.5) we derive

$$ [\partial_{x^{-}}\phi(x^{+},x^{-},{\bar x}),\phi(0)]\vert_{x^{+}=0}=
-{i\over 2}\delta^{2}(\bar x)\delta(x^{-}).\,\eqno(1.6)$$

\nl  Comparing (1.6) with  the equal-time commutator
$[\pi(x^{0},\vec x),\phi(0)]
\vert _{x^{0}=0}=-i\delta^{3}(\vec x)$, where $\,\pi=\partial_{t}\phi\,$,
it is suggested that the canonical
momentum in the light-front quantized
theory is $\;\sim \partial_{x^{-}}\phi $ and is thus not an independent
variable, a result we rederive below.

We note also that in the {\it front form} the Green's function are ordered with
respect to light-front time $x^{+}$ rather than $x^{0}\equiv t$.
However, in view of the microcausality
requirement, the retarded commutators $[A(x),B(0)]\theta(x^{0})$ and
$[A(x),B(0)]\theta(x^{+})$ do agree. In the regions $x^{0}>0, x^{+}<0$ and
$x^{0}<0, x^{+}>0$ where they seem to disagree $x^{2}$ is space like and they
are both vanishing. Such retarded commutators appear in the LSZ
reduction formulas [13] of the S-matrix  elements in
terms of the field operators.

\vskip 0.5cm

We describe briefly the elements of the {\it Poincar\'e algebra} on the
light-front and some of their properties.
In the system of coordinates $\,(x^{0},x^{1},x^{2},x^{3})\,$ with
the metric $g_{\mu\nu}=diag\,(1,-1,-1,-1)$ the generators of the
Poincar\'e algebra satisfy the following commutation relations:

$$[M_{\mu\nu},P_{\sigma}]=-i(P_{\mu}g_{\nu\sigma}-P_{\nu}g_{\mu\sigma}),$$

$$[M_{\mu\nu},M_{\rho\sigma}]=i(M_{\mu\rho}g_{\nu\sigma}+
M_{\nu\sigma}g_{\mu\rho}-M_{\nu\rho}g_{\mu\sigma}-M_{\mu\sigma}g_{\nu\rho})
\eqno(1.7)$$

\nl Introducing the convenient variables $J_{i}=-(1/2)\epsilon_{ikl}M^{kl}$
and $K_{i}=M_{0i}$ where $i,j,k,l=1,2,3$ we find

$$[J_{i},F_{j}]=i\epsilon_{ijk}F_{k}
\qquad\qquad for \qquad F_{l}=J_{l},P_{l} \,or \,K_{l}\,\eqno(1.8)$$

\nl while

$$[K_{i},K_{j}]=-i \epsilon_{ijk}K_{k}, \quad [K_{i},P_{l}]=-iP_{0}g_{ik},
\quad [K_{i},P_{0}]=iP_{i}, \quad [J_{i},P_{0}]=0.\eqno(1.9)$$

\nl We note that the six (kinematical)
generators $\,P_{l}, M_{kl}\,$ leave the hyperplane $x^{0}=0$ invariant.
In the light-front coordinates on the other hand there are seven such
generators
$\,P_{-},P_{1},P_{2}\,$, $M_{12}=-J_{3},\,M_{+-}=-K_{3},\,M_{1-}=
-(K_{1}+J_{2})/{\sqrt 2}\,
\equiv {-B_{1}}, \; M_{2-}=-(K_{2}-J_{1})/{\sqrt 2}
\equiv {-B_{2}}\;$ which leave the light-front $x^{0}+x^{3}=0$ invariant.
The remaining three generators related to the dynamic [1] are $P_{+},\,
M_{1+}=(K_{1}-J_{2})/{\sqrt 2}\equiv S_{1},\,
M_{2+}=-(K_{2}+J_{1})/{\sqrt 2}\equiv -S_{2}\,$. The generators $P^{+},P^{1},
P^{2},S_{1},S_{2},$ and $J_{3}$ commute with $P^{-}$. The generators
 $B_{1},B_{2},J_{3}$, which also commute with $P^{+}$,
 span the $E_{2}$ subalgebra of the 'little' group which leaves the light like
 vector $(n^{0}=1,0,0,n^{3}=1)$ invariant. The $K_{3}$ boosts along the
 3-direction ($[K_{3},P^{\pm}]={\mp}iP^{\pm}$)

 $$e^{-i\eta K_{3}}\,P^{\pm}\,e^{i\eta K_{3}}= e^{\mp \eta}\,
 P^{\pm},\,\eqno(1.10)$$

 \nl while under Galilean (transverse) boosts

$$ e^{-i \bar u.\bar B} P_{j} e^{i \bar u.\bar B}=
 (P_{j}+u_{j}P^{+}),\;  \qquad \quad  j=1,2\;\;and \;\;\; {\bar u}.{\bar B}=
 [u_{1}B_{1}+u_{2}B_{2}]\, \eqno(1.11)$$

 $$e^{-i\bar u.\bar B}\,P^{-}\,e^{i\bar u.\bar B}=
 P^{-}+\bar u.\bar P + {1\over 2}{\bar u}^{2}P^{+},
 \qquad e^{-i\bar u.\bar B}\,P^{+}\,e^{i\bar u.\bar B}= P^{+}.\eqno(1.12)$$

 \nl They are useful for constructing an eigenstates $\vert p^{+},p^{1},
 p^{2}\rangle $ starting, say, from the state described in the rest frame.
We collect also

$$[B_{1},P_{1}]=[B_{2},P_{2}]=i P^{+},\quad [B_{1},P_{2}]=[B_{2},P_{1}]=
0, \quad [B_{j},P^{-}]=iP_{j},\quad  [B_{j},P^{+}]=0\eqno(1.12)$$

$$[S_{j},P^{-}]=0,\qquad [S_{j},P^{+}]=iP_{j},\qquad [S_{1},S_{2}]=0,
\qquad [S_{1},P_{1}]=[S_{2},P_{2}]=iP^{-}, \eqno(1.13)$$

$$[B_{1},B_{2}]=0, \quad [B_{1},J_{3}]=-iB_{2},
\qquad [B_{2},J_{3}]=iB_{1}, \qquad [B_{j},K_{3}]=iB_{j}, \quad
[J_{3},K_{3}]=0.\eqno(1.14)$$

\vfill\eject

\nl {\bf 2- Spontaneous symmetry breaking mechanism in
light-front quantized scalar field theory:}
\medskip

 The  quantization of the scalar field theory in the {\it instant form} is
found in
the text books but the quantization on the light-front has been clarified
only recently. Working directly in the continuum
it was demonstrated [9,10] that corresponding to the local Hamiltonian in the
{\it instant formulation} we in fact now obtain a nonlocal Hamiltonian
in the {\it front form} formulation.
The nonlocality arises along the longitudinal direction $x^{-}$ because
of a new ingredient in the form of the
nonlocal {\it constraint equations} in the Hamiltonian formulation.
The treatment of the theory in the
usually adopted discretized formulation (assuming the finite volume) [14-18,9]
introduces spurious finite size contributions which
make the physical interpretations difficult. The infinite volume limit
may, however, be taken [9] which coincides with the continuum
formulation [10]. As argued in Sec. 1 such a nonlocality is not
unexpected and we pay the due
price for working with a simple vacuum on the light front. The
constraint equations allow  to describe
the tree level spontaneous symmetry breaking [10,14-18] and
suggest the modifications that would
be introduced by the quantum corrections [9]. The problem had been a
a challenge for a long time [8]. The reason seems to be that of not
distinguishing clearly between the bosonic condensate associated
with the scalar field and
the field which describes the (quantum) fluctuations (see also 2{\it (c)}).
We remind that the scalar field theory plays an important role in many
branches of physics. It is relevant, say,  in connection with the
generalized Ising models [19] in
the condensed matter theory, is an indispensable ingredient (Higgs sector)
of the Standard Model of electro-weak interactions, plays an important role
in describing inflationary cosmology, and in the construction, say, of
the heterotic string theory [6].

Dirac [1] discussed the problem of fitting together in a dynamical
theory the two general requirements: {\sl it should be quantized
and also incorporate the special
theory of relativity (ignoring gravitation)}.
He showed that Hamiltonian formulation, where we introduce a new element in the
form of Poisson brackets, is
a convenient procedure to attain this objective. We would apply  the
Dirac method [11] to construct this formulation.

\bigskip

\nl {\bf Discrete symmetry in two dimensions:}
\medskip

\nl {\it (a)- Continuum formulation:}
\medskip

In order to emphasize the new features which distinguish the {\it front form}
from
the {\it instant form} we consider first the simpler case of a
real massive scalar field theory in two dimensions.
The Lagrangian density in the {\it front form} is given by
$\, [{\dot\phi}{\phi^\prime}-V(\phi)],\,$ where an overdot and
a prime indicate the partial derivatives with respect to the light-front
time $\,\tau\equiv x^{+}=(x^0+x^1)/{\sqrt2}\,$
and the longitudinal coordinate
$\,x \equiv x^{-}=(x^0-x^1)/{\sqrt2}\,$ respectively. In contrast
to the case in the {\it instant form} dynamics
the $\dot\phi$ now occurs linearly in the Lagrangian and the eq. of motion
is $\,2\dot{\phi'}= -V'(\phi)\;$, where a prime on $V$ indicates
the variational derivative with respect to $\phi$. It
shows that the classical solutions, $\phi=const.\equiv
\omega\,$, are allowed and given by
solving $V'(\phi)=0$.  On the other hand the vacuum expectation value (vev)
$\,\langle vac \vert\phi\vert vac \rangle\,$ of the
quantized field must be a constant in view of the
requirement of the translation invariance of the theory. It is suggestive
that there is a certain  correspondence between the value $\omega$
with the tree level {\it vev} of the quantized field $\phi$.
We now note that if we integrate the
eq. of motion over $\,-L/2\le x\le L/2\,$, where $L\to\infty$,
we are led to the constraint equation, $\int dx\,V'(\phi)=
-2\,\partial_{\tau}[{\cal C}(\tau,L)]\,$, where $\,{\cal C}(\tau,L)=
[\,\phi(\tau,x=L/2)-\phi(\tau,x=-L/2)\,]\,$. The constraint thus seems
to depend on the boundary conditions imposed on $\phi$ and thus needs
clarification.
We should, however, first formulate the {\it physical problem}
at hand more precisely. Moreover, we need to write the equations
in Hamiltonian form as stressed above.
Taking into consideration the discussion made here we
propose to make the separation
$\;\phi(x,\tau)=\omega(\tau)+\varphi(x,\tau)\;$ for any fixed value
of light-front time $\tau$. The variable
$\omega\,$ corresponds to the  bosonic condensate or the
background field
while $\,\varphi\,$ describes the (quantum) fluctuations above the
condensate. This separation should be done independently
of whether we work in the continuum formulation or in the
 discretized one where $L$ is taken finite (finite volume),
 frequently employed. It should be emphasized, however, that a well defined
infinite volume limit must  exist if the theory is physical and
self-consistent [20].
At the classical level the $\varphi$ field
is an {\it ordinary} function of $x$ such that
($\int dx\,\vert\varphi\vert <\infty \,$)
and such that its Fourier  transform (or Fourier series)
along with its inverse are defined while
$\phi$ is a generalized function of $x$ due to the constant term $\omega$.
Since our primary interest is to discuss the vacuum states we will ignore
presently the $\tau$ dependence of $\omega$ and  return to the general case
latter below.

The Lagrangian then reads as

$$\int_{-L/2}^{L/2} dx\;[{\dot\varphi}\,{\varphi'}
-V(\phi)]\,,\eqno(2.1)$$

\noindent where for illustration purposes we take,
$\,V(\phi)=-(1/2)m^2 {\phi^2}
+({\lambda/4})\phi^{4}+const.\,$,  $\lambda> 0$, with the wrong sign for the
mass term and $L\to\infty$.
We next construct
the classical Hamiltonian formulation for the system
which may be used to make transition [1] to a relativistic and quantized theory
through the correspondence principle, in analogy with what we do in the
construction of the quantum mechanics or alternatively, by employing
the functional integral technique. The
canonical light-front Hamiltonian is easily
obtained to be $\int dx V(\phi)$. However,
in the {\it front form} dynamics there is no physical argument available
to minimize the light-front energy in order
to obtain the (classical) ground states. In equal-time formulation
we {\it add} new ingredients to the theory invoking physical
considerations such as
$\partial_{t}\varphi=0$ and $\partial_{x^{1}}\varphi=0$ which reduce the
energy and argue then to minimize the energy functional in order to obtain the
ground states. We will demonstrate below that {\it the light-front dynamics
already incorporates such information  in the theory through
self-consistency requirements and the new ingredients found in the form of the
constraint equations. The physical results in the two forms of dynamics
coincide even though obtained through different mechanisms}.
We remind that in the Hamiltonian approach for nonsingular Lagrangians
the number of dynamical
variables to describe the theory is doubled and it also contains
the first-order Hamilton eqs. whose number is twice
as compared to the number of the second-order
Lagrange eqs. The approach also introduces
a new object in the form of a Poisson bracket which allows us to
satisfy in the theory both
the requirements of the special theory of relativity and of the
transition to the quantum theory. It is also richer than the
Lagrangian formulation in that it allows for a broader set of general
transformations (on the phase space).
In a self-consistent Hamiltonian formulation the Lagrangian
formulation must be recovered [11].
In contrast to the {\it instant form} the light-front Lagrangian (2.1)
is singular and the Hamiltonian here
determines the  evolution of the dynamical system with changing
$\tau $ in place of $t$. We follow the Dirac method to build
the canonical framework at  a given $\tau$.
Indicating by $\pi(x,\tau)$ the momenta conjugate to $\varphi(x,\tau) $,
the primary constraint is found to be
$\,\Phi\equiv \pi-\varphi^\prime\approx 0\,$ while the canonical Hamiltonian
density is  $\,{\cal H}_{c}= V(\phi)\,$, with the symbol
${\approx}$ standing for the weak equality [11].
We postulate now the standard Poisson brackets at equal-$\tau$,
with the nonvanishing brackets satisfying,
$\{\pi(x,\tau),\varphi(y,\tau)\}=-\delta(x-y)$,
and assume for the preliminary Hamiltonian the expression

$$H^\prime(\tau) = {H_c}(\tau)
                        + \int dy\; u(\tau,y)\Phi(\tau,y),\eqno(2.2)$$

\noindent where  $u$ is a Lagrange multiplier function.
Using $\,\dot f=\{f,H'\}+{\partial f/\partial \tau}\,$ we find

$${\dot\Phi}\;=\;\{\Phi,H^\prime\}\;{\approx}\;
-{\, V'(\phi)}\,-\,2u^\prime.\eqno(2.3)$$

\noindent The persistency requirement $\;\dot\Phi\approx 0\;$
then results in a consistency condition involving the multiplier $u$
and does not generate a new constraint.
The only constraint  $\,\Phi(x)\approx 0\,$ in the theory
is second class [11] by itself since

$$\;\{\Phi(x),\Phi(y)\,\}=\,
-2{\partial_x} \delta(x-y)\;\equiv C(x,y)=
-C(y,x)\eqno(2.4)$$

\nl is nonvanishing. Its (unique) inverse with the correct symmetry property
is $C^{-1}(x,y)=-C^{-1}(y,x)=\,-{\epsilon(x-y)}/4$.
The equal-$\tau$ {\it Dirac bracket} $\,\{,\}_D\,$ is then constructed as

$$\{f(x),g(y)\}_D=\{f(x),g(y)\} + {1\over4}\int\int dudv \{f(x),\Phi(u)\}
\epsilon(u-v)\{\Phi(v),g(y)\}.\eqno(2.5)$$

\nl where we suppress $\tau $ for convenience of writing.
We verify that $\,\{f,\Phi\}_{D}=0\,$ for any arbitrary
functional $f(\varphi,\pi)$ and thereby we are allowed to set
 $\;\pi=\varphi^\prime\;$ even inside the Dirac bracket, e.g., treat it
 as a strong equality. The eqs. of motion are now
 given by $\dot f\,=\,\{f,H_{c}\}_{D}
+{{\partial f}/ {\partial\tau}} \,$ since $H'(\tau)$ reduces to
$\,H_c(\tau)\equiv P^{-}\, $, and whose explicit form is

$$\eqalign{ P^{-}\,&\equiv\int_{-L/2}^{L/2} dx \,V(\phi) \cr
&=\int_{-L/2}^{L/2} dx \,\Bigl [\omega(\lambda\omega^2-m^2)\varphi+
{1\over 2}(3\lambda\omega^2-m^2)\varphi^2+
\lambda\omega\varphi^3+{\lambda\over 4}\varphi^4
+const.\Bigr ]\,}\eqno(2.6)$$

\nl where $L\to\infty $.

The variable $\,\pi\,$ is not an independent one unlike in the case
of the {\it instant formulation} and from (2.5) we derive

$$\; \{\varphi(x,\tau),\varphi(y,\tau)\}_D=-(1/4)\epsilon(x-y)\;,\eqno(2.7)$$

\noindent We make a brief digression on the transition to the quantized
theory. To each dynamical variable we associate an operator in the quantized
theory and make the correspondence $i\{f,g\}_{D}\to [f,g] $ where  $[f,g]$
indicates a commutator (or anticommutator) between the operators. Such
procedure is frequently adopted to obtain the quantum mechanics of a
particle in the Heisenberg formulation. For example, from (2.7) we find
$\; [\varphi(x,\tau),\varphi(y,\tau)]=-(i/4)\epsilon(x-y)\;$  for the
$\varphi$ commutator in two dimensions. This agrees with the one suggested
from the considerations on the Lehmann spectral representation in Sec. 1.
We note that the antisymmetry of the Dirac bracket imposes that we define
$\epsilon(0)=0$. We also note that unlike in the equal-time case here
the commutator does not vanish on the light-cone
for non-coincident points with $x\ne y $
and we derive on using $\partial_{x}\epsilon(x-y)=2
\delta(x-y)$

$$\; \{\varphi'(x,\tau),\varphi(y,\tau)\}_D=-(1/2)\delta(x-y)\;,\eqno(2.8)$$

\nl and

$$\; \{\varphi'(x,\tau),\varphi'(y,\tau)\}_D=
-(1/2)\partial_{y}\delta(x-y)\;,\eqno(2.9)$$

\nl It is easy to show that the translations in the space direction $x^{-} $
are generated by

$$P^{+}(\tau)= \int dx\, (\varphi'(x,\tau))^{2}\;,
\qquad\qquad \qquad \varphi'(x,\tau)=
\{\varphi(x,\tau),P^{+}(\tau)\}_{D}.\eqno(2.10)$$

\nl For example,
$\{V(\phi(x,\tau),P^{+}(\tau)\}_{D}=\,\partial_{x}V(\phi(x,\tau))$ and
from which it follows that

$$\{P^{-}(\tau),P^{+}(\tau)\}_{D}\,
=\,V(\phi({\infty},\tau))-V(\phi({-\infty},\tau)).\eqno(2.11)$$

\nl The right hand side must vanish in a relativistic invariant theory.


The Hamilton's eq. for $\varphi$ is found to be

$$\eqalign {\dot\varphi(x,\tau)&=\{\varphi(x,\tau),P^{-}(\tau)\}_{D}\cr
&= -{1\over 4}\,\int dy \,\epsilon(x-y)\,{\delta V\over {\delta \phi(y,\tau)}}
}\eqno(2.12) $$

\nl and we derive from it the Euler-Lagrange eq.

$$\dot\varphi'(x,\tau)=-{1\over 2}\,{\delta V\over {\delta \phi(x,\tau)}}
\,\eqno(2.13)$$

\nl The Hamiltonian formulation constructed above is thus
self-consistent with the Lagrange formulation. If we
substitute the value of $V'(\phi)$ obtained
from (2.13) into (2.12) we find after
an integration by parts

$$ \dot\varphi(x,\tau)=\dot\varphi(x,\tau)-
{1\over 2}\,{ \Bigl [\dot\varphi(\infty,\tau)\epsilon(\infty-x)
-\dot\varphi(-\infty,\tau)\epsilon(-\infty-x)\Bigr].}\, \eqno(2.14) $$

\nl Considering finite values of $x$ we must then require
$\,\dot\varphi(\infty,\tau)+\dot\varphi(-\infty,\tau)=0\,$. On the other hand
the equal-$\tau$ commutator of $\varphi$  obtained
above may be realized through the  momentum space expansion given
below in (2.16). Integrating over $x$ the expansion of $\varphi'(x,\tau)$
we easily show that $\,\varphi(\infty,\tau)-\varphi(-\infty,\tau)=0$.
Combining the two results we are led to
$\partial_{\tau}\varphi(\pm\infty,\tau)=0$ as a consistency condition.
This is similar to $\partial_{t}\varphi(x^{1}=\pm\infty,t)=0$ which is
imposed from the outside in the equal-time formulation based upon the
physical considerations.

On integrating (2.13) over the longitudinal space coordinate
it follows that we must (also) satisfy  the following
{\it constraint equation} in the light-front Hamiltonian framework

$$ \eqalign {\beta(\tau) &\equiv\,lim_{L\to\infty}
{1\over L}\int_{L/2}^{L/2} dx \,V'(\phi) \cr
&= \omega(\lambda\omega^2-m^2)+lim_{L\to\infty} {1\over L}
 \int_{-L/2}^{L/2} dx \Bigl[ \,(3\lambda\omega^2-m^2)\varphi +
 \lambda (3\omega\varphi^2+\varphi^3 ) \,
\Bigr]=0}\,\eqno(2.15)$$

\nl The Hamiltonian formulation in the present case thus contains a new
ingredient in the form of a nonlocal constraint (2.15).
This is not unexpected in view of the discussion give in Sec. 1.
In ref. 1 we find some illustrations where restrictions on the
potential arise due to the necessity of incorporating special
relativity in the theory. Because of the constraint eq. the light-front
Hamiltonian (2.6) is nonlocal and much
involved compared with the local polynomial
form for interaction  in the corresponding equal-time formulation.

At the classical (or tree) level the integrals appearing in
(2.15) are convergent (since $\int dx \vert\varphi\vert <\infty$). In
the continuum limit, when  $L\to\infty$, we find the result $V'(\omega)=0$
which determines the tree level values for the condensate $\omega$ and they
are the same as those found in the {\it instant formulation},
where the condition is added to the theory (imposed)
by appealing to physical considerations for
minimizing the energy functional. Similar comments hold true as regards
$\partial_{\tau}\varphi(\pm\infty,\tau)=0\,$.  Such considerations
in general are {\it not}
available in the {\it front form} dynamics. In its
place constraint equations like (2.15) or consistency conditions like
(2.14) arise in the theory itself. The two forms of dynamics should clearly
lead to the same physical results even though attained by different
paths. This  will be explicitly seen  in the description of the
spontaneous symmetry breaking considered below.


In the {\it quantized theory} the field $\varphi$ and its commutation
relations are obtained through the correspondence
$i\{f,g\}_{D}\to [f,g]$. It is easy to verify that the equal-$\tau$
$\varphi$ field
commutator,
$\; [\varphi(x,\tau),\varphi(0,\tau)]=-(i/4)\epsilon(x)$, can be realized
in the momentum space through the following
(momentum space) expansion of the field (suppressing $\tau$)

$$\varphi(x)= {1\over {\sqrt{2\pi}}}\int_{-\infty}^{\infty} dk\;
{\theta(k)\over {\sqrt{2k}}}\;
[a(k)e^{-ikx}+{a^{\dag}}(k)e^{ikx}]\,\eqno(2.16)$$

\noindent where $a(k)$ and ${a^{\dag}}(k)$
satisfy the canonical equal-$\tau$ commutation relations,
$[a(k),{a(k^\prime)}^{\dag}]=\delta(k-k^\prime)$,
$[a(k),a(k^\prime)]=0$, and $[{a(k)}^{\dag},{a(k^\prime)}^{\dag}]=0$.
We note the  presence of $\,\theta(k)\,$ in (2.16) originating as a
consequence of the integral representation of the sgn function,
$\,\epsilon(x)\,=(i/\pi){\cal P}\int (dk/k) e^{-ikx}$.
Also, in contrast to the case of
equal-time formulation, there is no dependence on mass in (2.16).

The vacuum state is defined
by  $\,a(k){\vert vac\rangle}=0\,$,
$k> 0$.
The longitudinal momentum operator and the light-front energy operators are
$\int dx:\varphi'{^2}: $ and $P^{-}=H=\int dx :V(\phi): $ respectively. Here
we normal order with
respect to the creation and destruction operators to drop
unphysical infinities.
We find $\,[a(k),P^{+}]=k\,a(k)\,$,
$\,[{a^{\dag}}(k),P^{+}]=-k\,{a^{\dag}}(k)\,$.  The tree level
description of the spontaneous (discrete) symmetry breaking may be given as
follows. The values of $\omega=
\,{\langle\vert \phi\vert\rangle}_{vac}\;$ obtained from the tree level
condition $V'(\omega)=0$
{\it characterize} the possible  vacua of different type in the theory.
Distinct  Fock spaces corresponding to different values of $\omega$
are built as usual by applying the creation operators on the corresponding
vacuum state.  The $\omega=0$ corresponds to a {\it symmetric phase}
since the Hamiltonian is then found symmetric under the discrete symmetry
transformation $\varphi\to -\varphi$. For $\omega\ne0$ this symmetry is
clearly violated and the  system is
in a {\it broken or asymmetric phase}.  In the case of the
right sign for the mass
term, $m^2\to -m^2$,  and $\lambda=0$
the system is found in the symmetric phase at the tree level.
When the interaction is switched on, the constraint eq. (2.15) shows,
however, that the
symmetric phase may become unstable due to the quantum corrections and
the system may undergo a phase transition, as the coupling
constant increases, to the broken phase. For the wrong sign for
the mass term we have the possibility of both types of phases
already at the tree level. It should be stressed that we do {\it not} have
any physical arguments, like for $P^{\pm}$,
in the {\it front form} theory to normal order the constraint
equation (2.15) and consequently the tree level value of the condensate does
obtain high order quantum corrections as dictated by the renormalized
constraint equation. A factor $L$ may arise in the numerator which
cancels the $L$ in the denominator in the integrals involved in (2.15).
A self-consistent {\it front form}
Hamiltonian formulation can thus be built in the
continuum which also can describe the ssb.
The phase transition in two dimensions can be
described (Sec. 3) in the renormalized theory based on
$P^{-}$, $\beta=0$, and the light-front commutator.

\bigskip

\nl {\it (b)- Discretized formulation in finite volume:}
\medskip

It is instructive to rederive the continuum theory above
as the  infinite volume limit of the {\it discretized formulation} in the
finite volume [9]. It is some times wrongly
affirmed in the literature that the infinite volume limit of the
discretized formulation does not exist. We make  the
Fourier series expansion of the field $\varphi\,$ and write

$$\phi(\tau,x) =\omega+{{q_{0}(\tau)}\over\sqrt{L}}+
{1\over\sqrt{L}}\;{{\sum}'_{n}}\;
{q}_n(\tau)\;e^{-ik_n x}\equiv \omega +\varphi(\tau,x)\,\eqno(2.17)$$

\noindent where the periodic boundary conditions are
assumed for convenience with $\,\Delta=(2\pi/L)$,$\,k_n=n\Delta $,
$\,n=0,\pm 1,\pm 2, ..\,$, and $L$ is now finite.
The discretized Lagrangian obtained by integrating the
Lagrangian density in (2.1) over the finite interval $\;-L/2 \le x \le
L/2\; $  is given by

$$\;i{{\sum}_{n}}\;k_n \,{q}_{-n}\;{{\dot q}}_{n}\;-
                                  \int^{L/2}_{-L/2} dx \; V(\phi)\eqno(2.18)$$

\noindent The momenta conjugate to ${q}_n$ are then
 ${p}_n=ik_n{q}_{-n}$
 and the canonical Hamiltonian is obtained from the expression in (2.6).
The primary constraints are thus
${p}_0 \approx 0\;$ and $\;{\Phi}_n \equiv
\;{p}_n-ik_n{q}_{-n}\approx 0\;$ for $\,n\ne 0\,$.
We postulate initially the standard Poisson brackets at equal
$\tau\,$, viz, $\{{p}_m,{q}_n\}=-\delta_{mn}\;$
and define the preliminary Hamiltonian

$$H'=H_c+{{\sum}'_{n}}\;{u}_n{\Phi}_n
+\,{u}_0 \,{p}_0\,. \eqno(2.19)$$

\noindent On requiring the persistency in $\,\tau\,$ of these constraints
we find the following weak equality relation

$${{\dot p}}_0\,=\,\{{p}_0,\,H'\}\,=-{\partial H'\over \partial
q_{0}}\,=
\,-{1\over\sqrt{L}}
\int^{L/2}_{-L/2}\,dx\,V'(\phi)\;\equiv \,-{1\over\sqrt{L}}\;\beta(\tau)\,
 \approx 0\, ,\eqno(2.20)$$

\nl and for $n\ne0$

$${\dot\Phi}_n\;=\;\{\Phi_n,H'\}\,=\,-2i{{\sum}'_{n}}k_n\,{u}_{-n}\,-\,
{1\over\sqrt{L}}
\int^{L/2}_{-L/2}\;dx\;V'(\phi)\;e^{-ik_n x} \approx 0\,.\eqno(2.21)$$

\noindent From (2.20) we obtain an interaction dependent secondary constraint
$\;\beta\approx 0$ while (2.21)
is a consistency requirement for determining ${u}_n,\,n\ne 0\,$.
Next we extend the Hamiltonian to

$$H''\;=\,H'+ \nu(\tau)\beta(\tau),\eqno(2.22)$$

\noindent and check again the persistency of all the constraints
encountered above making use of $H''$. We check that no more secondary
constraints are generated if we set $\nu\approx 0$ and we are left only with
consistency requirements for determining the multipliers
$\,u_{n}\,, u_{0}$.

We verify that all the constraints $\,p_{0}\approx 0$, $\,\beta
\approx 0$, and $\,\Phi_{n}\approx 0\,$ for $\,n\ne 0\,$
are second class. They may be implemented in
the theory by defining Dirac brackets
and this may be performed iteratively. We find $(\,n,m\ne 0\,)$

$$\{{\Phi}_n, {p}_0\}\,=\,0,\qquad\qquad
\{{\Phi}_n,{\Phi}_m\}\,=\,-2ik_n \delta_{m+n,0}\;,
\eqno(2.23)$$

$$\{{\Phi}_n,\beta\}\,=\,\{{p}_n,\beta\}\,
=\,-{1\over \sqrt{L}}\int^{L/2}_{-L/2}dx\;\lbrack \,V''(\phi)-V''({\omega +
q_{0}\over
\sqrt{L}})\,\rbrack \,e^{-ik_nx}\,
\equiv \,-{{\alpha}_{n}\over\sqrt{L}},\,\eqno(2.24)$$

$$\{{p}_0,\beta\}\,=\,-{1\over\sqrt{L}}\int^{L/2}_{-L/2}dx\;V''(\phi)\,
\equiv \,-{\alpha\over\sqrt{L}},\,\eqno(2.25)$$

$$\{{p}_0,{p}_0\}\,=\,\{\beta,\beta\,\}\,=\,0\,.\eqno(2.26)$$

\nl The explicit expressions of $\alpha_{n}$ and $\alpha$ appear
below in the
numerator and the denominator of eq. (2.29).

We implement first the pair of constraints $\;{p}_0\,\approx 0,
\,\beta\,\approx 0$. The Dirac bracket $\{\}^*$
with respect to them is easily constructed

$$\{f,g\}^*\,=\{f,g\}\,-\lbrack\,\{f,{p}_0\}\;
\{\beta,g\}-\,(p_0\,\leftrightarrow
\beta)\rbrack\,({\alpha\over\sqrt{L}})^{-1}.
\eqno(2.27)$$

\noindent We may then set ${p}_0\,=0$
and $\beta \,=0$ as strong relations.
and the variable $\,{p}_0\,$ is thus
removed from the theory. We conclude easily by inspection that the brackets
$\{\}^*\,$ of the surviving canonical variables coincide with the standard
Poisson brackets except for the ones involving $q_{0}$ and
$\,p_{n}\,$ ($n\ne0$)

$$\,\{q_0,{p}_n\}^*\,=
\{q_0,{\Phi}_n\}^*\,=-({\alpha^{-1}}{\alpha}_n)\,\eqno(2.28)$$

\noindent For the explicit expression of the potential given above
we find  $\,\{q_0,{p}_n\}^*\,=$

$$-{
\,{3\lambda\,\lbrack\,2(\omega+q_{0}/{\sqrt L})\,{\sqrt L}
 q_{-n}+\,\int^{L/2}_{-L/2}\,dx\,\varphi^{2}
\,e^{-ik_{n}x}\,\rbrack}
\over {\lbrack\,3\lambda\,({\omega+q_{0}/\sqrt{L}})^{2}-m^{2}\,\rbrack\,L\,+
6\lambda(\omega+q_{0}/{\sqrt L})\int_{L/2}^{L/2} dx \varphi +
\,3\lambda \,\int^{L/2}_{-L/2}\,dx\,\varphi^{2}}}\,\eqno(2.29)$$

Next we implement the remaining constraints $\,\Phi_{n}\approx 0\,$
($\,n\ne 0$). We have

$$ C_{nm}\,=\,\{\Phi_{n},\Phi_{m}\}^*\,=
\,-2ik_{n}\delta_{n+m,0}\,\eqno(2.30)$$

\noindent and its inverse is given by  $\,{C^{-1}}_{nm}\,=\,(1/{2ik_{n}})
\delta_{n+m,0}\;$. The {\it final} Dirac bracket which takes care of all
the constraints of the theory is then given by

$$\{f,g\}_{D}\,= \,\{f,g\}^{*}\,-\,{{\sum}'_{n}}\,{1\over {2ik_{n}}}
\{f,\,\Phi_{n}\}^*\,\{\Phi_{-n},\,g\}^*.\,\eqno(2.31)$$

\noindent Inside this final bracket all the constraints may be treated
as strong relations and we may now in addition write $\,p_{n}\,=\,ik_{n}
q_{-n}\,$. It is straightforward to show that

$$\{q_0,{q}_0\}_{D}\,=0,\qquad
\{q_0,{p}_n\}_{D}\,=\{q_0,\,ik_{n}
q_{-n}\}_{D}\,={1\over 2}\,\{q_0,{p}_n\}^*,
\qquad\{q_{n},p_{m}\}_{D}\,={1\over 2}\delta_{nm}.\eqno(2.32)$$




In order to remove the {\it spurious finite volume effects} in discretized
formulation we must take the {\it continuum limit} $\,L\to\infty\,$ [20].
We follow the well known procedure:
$\Delta=2\,({\pi/{L}})\to dk\,,\,k_{n}=n\Delta\to k\,,\,\sqrt{L}\,
q_{-n}\to\,lim_{L\to\infty}
 \int_{-L/2}^{L/2}{dx}\,\varphi(x)\,e^{ik_{n}x}\equiv\,\int_{-\infty}^
{\infty}\,dx\, \varphi(x)\,e^{ikx}\,=\,\sqrt{2\pi}{\tilde\varphi(k)}$
for all $\,n $,
$\,\sqrt{2\pi}
\varphi(x)\,=\int_{-\infty}^{\infty}\,dk\,\tilde\varphi(k)\,e^{-ikx}\;$,
and  $\,lim_{L\to\infty}(q_{0}/{\sqrt{L}})= 0\,$.
{}From $\,\{\sqrt{L}q_{m},\sqrt{L}q_{-n}\}_{D}\,=\,
L\,\delta_{nm}/({2ik_{n}})\,$ following from the Dirac bracket between
$q_{m}$ and $p_{n}$ for $n,m\ne 0$ in (2.32)
we derive, on using  $\,L\delta_{nm}\to lim_{L\to\infty}\int_{-L/2}^{L/2}dx
e^{i(k_{n}-k_{m})x}\,= \int_{-\infty}^{\infty}dx e^{i(k-k')x}=
\,{2\pi\delta(k-k')}$, that

$$\{\tilde\varphi(k),\tilde\varphi(-k')\}_{D}\,={1\over {2ik}}\,\delta
(k-k')\,\eqno(2.33)$$

\noindent where  $ k,k'\,\ne 0$.
On making use of the integral representation of the sgn
function, $\,\epsilon(x)\,=(i/\pi){\cal P}\int_{-\infty}^{\infty}$ $(dk/k)
e^{-ikx}\,$ we are led from (2.33) to the light-front Dirac bracket (2.7) for
$\varphi$ obtained in  the continuum formulation above.

\noindent From (2.29) and (2.32) we derive

$$\{{q_{0}\over{\sqrt L}},\varphi'(x)\}_{D}\,=-({3\lambda\over 2})\,
{{\,\lbrack\, 2\omega\varphi(x)+
\varphi^{2}(x)\rbrack}\over {[(3\lambda\omega^{2}-m^{2})L\,+
6\lambda\int dx \, \varphi+\,3\lambda\int dx\,\varphi(x)^{2}]}}\,\eqno(2.34)$$

\noindent where $L\to\infty$. The eq. (2.34) is consistent in the
continuum limit
for the values of $\omega$ given at the tree level by
$V'(\omega)=0$. The constraint eq. (2.20), $\,\beta=0\,$, in the discretized
formulation,
goes over to the expression (2.15) of the continuum formulation and
the eqs. (2.12) and (2.13) are also recovered in the infinite volume limit.

It is interesting to recount {\it the history of the constraint equation}
 in the light-front framework. In the decade of 1970 the light-front
commutator was reobtained [21] by other methods not following the Dirac
procedure and the constraint escaped the observation. In 1976 the procedure
was attempted [14] in two dimensional scalar field theory using the
discretized formulation. The
constraint $p_{0}\approx 0$ was, however,  missed. The constraint equation
was noted but its implications ignored. Latter in 1989 it was remarked [16],
again in the context of the discretized formulation,
that the light-front quantization of the scalar field theory would be
extremely difficult due to the complexity of the constraint eq. which relates
the zero mode $q_{0}$ with nonzero modes $q_{n}$ with $n\ne0$ (see (2.15)).
Moreover, at the quantized level the zero mode (in the finite volume) is an
operator which does not commute with the nonzero modes (see (2.34)), making
it necessary to order it in the constraint itself.
Here also the Dirac procedure was not followed and arguments
were based essentially on the equation of motion. In 1991 it was
proposed [17] to modify the Dirac method introducing $p_{0}\approx 0$
from outside; this may not be regarded as gauge-fixing constraint since we
do not have any first class constraint in the theory considered.
There was
also some confusion caused by not distinguishing between the zero
mode of $\varphi$ and the bosonic condensate $\omega$, clarified
only latter. Since 1985, with the proposal [8] of  DLCQ-{\sl Discretized
light cone quantized theory } in the context of
perturbation theory, the zero mode (and the condensate) were
ignored  until
recently. The Dirac quantization of the light-front scalar field theory
directly in the continuum while separating the condensate [10,9]
was considered
only towards the end of 1991. It was strongly believed
since 1977, when it was perhaps first mentioned [15],
that it was not possible to
take the infinite volume limit of the discretized formulation just discussed,
ignoring
strangely enough the conflict with a basic principle [20].
It was sometimes also affirmed
that the light-front quantized field theory
could not even  be construct directly in the continuum. From our discussion
we conclude on the contrary. The theory seems manageable only in the continuum
formulation and there are  no signs of any inconsistency when properly
constructed following the Dirac method without modifications. We loose
control over the self-consistency checks if any modifications be introduced in
the procedure and even the doubts would arise on the genuinity of the
constraint equation obtained. The theory constructed above can also
describe the spontaneous symmetry breaking of both the discrete as well as
the continuous symmetry (see below) and  it contains  both the tree
(classical) as well as quantized level descriptions.
The {\sl suggestion that the Hamiltonian in
the light-front context may be nonlocal does not occur in
any of the earlier works}.
\bigskip
\nl {\it (c)- Continuum formulation with $\omega$ dynamical:}
\medskip
In case we maintain [10] the $\tau$ dependence in $\omega$ the Lagrangian is

$${\cal C}\dot\omega+\int_{-L/2}^{L/2} dx\,[{\dot\varphi}\,{\varphi'}
-V(\phi)]\,,\eqno(2.35)$$

\noindent where
$\,{\cal C}(\tau)=
[\,\varphi(\tau,x=L/2)-\varphi(\tau,x=-L/2)\,]\,$
and  $L\to\infty$.
The eqs. of motion are

$$\eqalign{\dot \varphi'& =(-1/2)\,V'(\phi),\cr
\dot {\cal C }&= -\int_{-L/2}^{L/2} dx  \,V'(\phi)\,}\eqno(2.36) $$

\nl Integrating the first on  $x$ for
($-L/2<x<L/2$) and comparing with the second we find
$\,\dot {\cal C}=0\,$ and consequently the constraint (2.15).
It is interesting to note that if we consider  $\,{\cal C}\,$ also
as a dynamical variable we obtain
$\dot \omega = 0 $ among the eqs. of motion.

In the case   $\varphi$ is an ordinary function admitting Fourier
transformation as well as its inverse the surface term
${\cal C}$ drops out since $\,\varphi(\pm\infty,\tau)=0\,$ (from the Fourier
transform theory). We note that the
Fourier transform of the generalized function $\phi$
is given by
$\,\tilde \phi(k,\tau)={\sqrt {2\pi}}\,\omega(\tau)\,\delta(k)+\,
\tilde\varphi(k,\tau)$. Indicating the canonical momenta  conjugate to
$\omega$ and  $\varphi$  by  $p$ e $\pi$ respectively, the
primary constraints are  $p\approx 0$ and  $\Phi\equiv \pi-\varphi'
\approx 0$ and we start from the preliminary  Hamiltonian

$$H^\prime(\tau) = {H_c}(\tau) + {\mu(\tau)}p(\tau)
                        + \int dy\; u(\tau,y)\Phi(\tau,y),\eqno(2.37)$$

\noindent where   $\mu$ e $u$ are Lagrange  multipliers. We find

$${\dot p}\;=\;\{p,H^\prime\}\;{\approx}\;-\int dx
{\; V^\prime(\phi)}\;
\equiv \;-\beta(\tau),\eqno(2.38)$$

$${\dot\Phi}\;=\;\{\Phi,H^\prime\}\;{\approx}\;
-{\, V'(\phi)}\,-\,2u^\prime.\eqno(2.39)$$

\noindent The  Dirac procedure the leads to the three second class
constraints $\;{p\approx0,\; \beta\approx0, \;\Phi\approx0}$.  In view of
$\;\;\{\beta(\tau),p(\tau)\}\equiv \alpha(\tau)=\int dx \;{V''(\phi)}\;$,
$\{\beta,\beta\}=\{p,p\}=0\;$, we define the modified bracket   $\,\{,\}^*\,$

$$\{f(x),g(y)\}^*=\{f(x),g(y)\}-{1\over{\alpha}}[\{f(x),p\}\{\beta,g(y)\}
-(\beta\leftrightarrow {p})]\,,\eqno(2.40)$$

\nl where

$$\,\alpha(\tau)=\int dx \,V''(\phi)=
L\,(3\lambda\omega^2-m^2)\,+6\lambda \omega \,\int dx
\,\varphi\,+3\lambda
\,\int\,dx\,{\varphi^2},\eqno(2.41)$$

\nl We verify that for the independent dynamical variables only
$\;\{\omega,\pi\}^*=\{\omega,\Phi\}^*=
\,-{\alpha^{-1}}\,{V''(\phi)}\;$ do not coincide with the corresponding
Poisson brackets $\{,\}$. Defining the final Dirac bracket $\{,\}_{D}$ by

$$\{f(x),g(y)\}_D=\{f(x),g(y)\}^* + {1\over4}\int\int dudv \{f(x),\Phi(u)\}^*
\epsilon(u-v)\{\Phi(v),g(y)\}^*.\eqno(2.42)$$

\nl we can also implement   $\Phi= 0$. From (2.42) we derive

$$\; \{\varphi(x),\varphi(y)\}_D=-(1/4)\epsilon(x-y)\;,\eqno(2.43)$$

$$\; \{\omega,\pi(x)\}_D=\,\{\omega,\varphi'(x)\}_D=\,
{1\over2}\{\omega,\pi(x)\}^{*},\qquad\qquad
\, \{\omega,\omega\}_D=0\,\eqno(2.44)$$

\nl The constraint  $\beta=0$ is interpreted as in the discusssion
given above. At the tree level $\alpha\to\infty$ which leads to
 $\,\{\omega,\pi\}^{*}\,=-{\alpha}^{-1}\, V''(\phi)\to 0\,$. It follows from
 (2.44) that  $\,\{\omega,\varphi(x)\}_D=0\,$, which leads to
 $\dot \omega=0$, in agreement with the constant values for $\omega$.
 We also verify that the  Lagrange eq. for  $\varphi$ is also recovered.
It is thus possible to construct a self-consistent Hamiltonian formulation
in the continuum with the proposal of separating first the condensate
from the fluctuations represented by $\varphi$.

\bigskip

\nl {\bf Continuous symmetry in $3+1$ dimensions:}
\medskip

The extension to $3+1$ dimensions and to global continuous symmetry case
is straightforward. Consider the multiplet of real scalar fields
$\phi_{a} (a=1,2,..N)\,$ with nonzero mass which transform as an isovector
under the global isospin transformations of the internal symmetry group
 $O(N)$. We separate the condensate variables and write
 $\phi_{a}(x,\bar x,\tau)
=\omega_{a}+\varphi_{a}(x,\bar x,\tau)$ while
$\omega_{a}$ is assumed independent of  $\tau$. The classical
Lagrangian density

$${\cal L}=[\;{\dot\varphi_{a}}{\varphi'_{a}}-
{(1/ 2)}(\partial_i\varphi_{a})(\partial_i\varphi_{a})-V(\phi)\;]
\,\eqno(2.45)$$

\nl is invariant with respect to the global $O(N)$ symmetry group. Here
$i=1,2$ indicate the transverse space directions. The eqs. of motion are
given by
$\,2\dot\varphi_{a}'=[-V_{a}'(\phi)+\partial_{i}\partial_{i}\varphi_{a}]\,$.
Following the Dirac procedure as above we find

$$\; [\varphi_{a}(x,\bar x,\tau),\varphi_{b}(y,\bar y,\tau)]=
-(i/4)\delta_{ab}\epsilon(x-y)\delta^{2}(\bar x-\bar y)\;,\eqno(2.46)$$

$$P^{-}(\tau)=\int\,dx d^2x \Bigl [V(\phi)+{1\over 2}
(\partial_{i}\varphi_{a})(\partial_{i}\varphi_{a})\Bigr],
\qquad\qquad\qquad P^{+}=\int\,dx d^2x \pi_{a}\varphi_{a},\,\eqno(2.47)$$

\nl where  $\pi_{a} (x,\bar x,\tau)=\varphi_{a}'(x,\bar x,\tau)$,
$\,\dot\varphi_{a}(\pm\infty,\bar x,\tau)=0$, and the set of coupled
constraint equations $\beta_{a}=0$ when expanded in Taylor series is given as

$$L\,V'_a(\omega)+
\,V''_{ab}(\omega)\int dx \varphi_{b}+\,{1\over {2!}}
V'''_{abc}(\omega)\int dx \varphi_b\varphi_c+...=0. \eqno(2.48)$$

\nl The momentum space expansion of the fields is now

$$\varphi_{b}={1\over ({{\sqrt 2}\pi})^{3}}\int dk d^{2}{\bar k}\;
{\theta(k)\over {\sqrt{2k}}}\;
[a_{b}(k,\bar k,\tau)e^{-i(kx+\bar k.\bar x)}+{a_{b}^{\dag}}(k,\bar k,\tau)
e^{i(kx+\bar k.\bar x)}]\,\eqno(2.49)$$

\nl where  $[a_{b}(k,\bar k,\tau),{a_{c}(k',{\bar k'},\tau)}^{\dag}]=
\delta_{bc}\delta(k-k')\delta^{2}({\bar k}-{\bar k'})\,$  etc.
At the tree level the values of  $\omega_{a}$ are obtained from
$V_{a}'(\omega)=0$ where  $V'_{a}$ indicates the variational derivative of
$V(\phi)=-(1/2)m^2\phi^2+
(\lambda/4)\phi^{4}+ const.\,$,
with respect to  $\phi_{a}$, with  $\phi^{2}=\phi_{a}\phi_{a}$.

The case of discrete symmetry in $3+1$ dimensions is obtained
here when there is only one real field.

\bigskip

Consider next the discussion of the field theory symmetry generators
and the description of the {\it spontaneous symmetry breaking of the
continuous symmetry}.
The classical theory is invariant
under global isospin rotations $\,\delta\varphi_{a}=-i
\epsilon_{\alpha}(t_{\alpha})_{ab}
\varphi_{b}\,$, $\,\delta\omega_{a}=-i\epsilon_{\alpha}(t_{\alpha})_{ab}
\omega_{b}$ where  $\alpha,\beta=1,2,..N(N-1)/2 $ are the group indices,
$t_{\alpha}$ are hermitian and antisymmetric generators of the group,
and  $\,[t_{\alpha},t_{\beta}\,]=\,
if_{\alpha\beta\gamma}\, t_{\gamma}$. The classical conserved
Noether  currents are given by
${J^{\mu}}_{\alpha}=-i{\partial^{\mu}\phi^{T}t_{\alpha}\phi}=
-i{\partial^{\mu}\varphi^{T}t_{\alpha}\varphi}-i(t_{\alpha}\omega)^{T}
\partial^{\mu}\varphi\,$. In the light-front quantized theory the
field theory symmetry generators are

$$\eqalign { \,G_{\alpha}(\tau)&=\int {d^{2}\bar x }dx \, J^{+}\cr
&=-i\int d^{2}{\bar x} dx[ {\varphi'_{a}}
(t_{\alpha})_{ab}\varphi_{b}-i(t_{\alpha}\omega)_{a}
\varphi'_{a}] \cr
&={-i\int d^{2} {\bar x} dx
{\varphi'_{a}(t_{\alpha})_{ab}\varphi_{b}}
=\,\int d^{2}{\bar k}\,dk \, \theta(k) {a_{a}(k,{\bar k})
^{\dag}}\,(t_{\alpha})_{ab}\,a_{b}(k,{\bar k})}\, }\,\eqno(2.50)$$

\nl in  view of $\,\varphi(\pm\infty,\bar x,\tau)=0\,$ and
where we used also  the (light-front) expansion (2.49) of the field.
The continuous symmetry generators come out already normal ordered
and as such they annihilate the vacuum state independent of the values
assumed by the condensates. We define the system to be in the
{\it symmetric phase} when all the $\omega_{a}$ are vanishing while it is
defined to be in the {\it broken or asymmetric phase} when some of these are
nonvanishing.

The situation in the conventional equal-time formulation is different.
The symmetry generators here are given by

$$\eqalign { Q_{\alpha}(x^{0})&=\int d^{3}x \, J^{0}\cr
&=-i\partial_{0}\varphi_{a}(t_{\alpha})_{ab}\varphi_{b}
-i(t_{\alpha}\omega)_{a}\int d^{3}x
{{d\varphi_{a}}\over dx_{0}}} \eqno(2.51)$$

\nl In the asymmetric phase the generators do not annihilate now the vacuum
state and the symmetry of the vacuum is broken. The generators, however, are
conserved even in  the quantized theory, since
$\,[Q_{\alpha},\phi_{a}]=-(t_{\alpha}\phi)_{a}$
e consequently, $\,[Q_{\alpha},H(t)]=0\,$. We call it the spontaneous
symmetry breaking because the Hamiltonian remains invariant under the
symmetry transformations but the vacuum state does not. In fact,
the first term  on the last line in (2.51)
annihilates the vacuum like in the earlier case
but the second one gives a vanishing contribution
only for the generators
for which  $\,(t_{\alpha}\omega)=0\,$. The set of such linearly independent
generators define the group
of residual symmetry (of the vacuum state) in the theory.

Returning to the light-front field theory case we verify that
$\,[G_{\alpha},\phi_{a}\,]=\,-(t_{\alpha})_{ab}\varphi_{b}-
(t_{\alpha}\omega)/2\, $,
$\,[G_{\alpha},\omega_{a}\,]=0\,$. They imply that in the asymmetric phase
the symmetry of the quantized theory Hamiltonian is broken while the symmetry
of the vacuum state is preserved. Only the generators
(or the linear combination of original generators)  for which
$\,(t_{\alpha}\omega)=0\,$ are left conserved, e.g., the symmetry
transformations associated with them leave the Hamiltonian invariant.
The set of such  generators give rise to the residual symmetry group of the
Hamiltonian operator and of the quantized theory.

The spontaneous symmetry breaking in the {\it front form} is
described as follows. At the tree level
a particular solution $(\omega_{1},\omega_{2},\omega_{3},...)$ of
$V_{a}'(\omega)=\,\omega_{a}(\lambda\omega^{2}-m^2)=0\,$ determines a
a fixed direction in the isospace
which {\it characterizes} a (non-perturbative)
vacuum state, $\;{\langle 0\vert \phi_a\vert 0\rangle}_{\omega}=
\omega_a $.
The Fock space of this
sector in the quantized theory is built by applying
the particle creation operators on the  vacuum state.
In the symmetric phase, both the vacuum and the Hamiltonian are invariant
under the internal symmetry group. In the asymmetric phase the vacuum
remains invariant under the initial symmetry group
but the Hamiltonian does not remain so under some of the symmetry
transformations. The residual symmetry group in both the {\it front form} and
the {\it instant form} is determined from the condition ($t_{\alpha}\omega)=0$.
The total number of the corresponding generators does not depend on the
particular choice of the isovector as long as $(\lambda\omega^{2}-m^2)=0\,$.
There is an infinite degeneracy corresponding to the continuum of
orientations of the (condensate or background field)
isovector in the isospin space satisfying this condition. This corresponds to
the infinite
degeneracy of the vacuum states in the equal-time case.
The number of Goldstone bosons may  also be counted [22]
and is the same as in the usual formulation.
Their number (ignoring the case of pseudo-Goldstone bosons)
is the difference in the number of generators of the
original and the residual symmetry group.
The values of the condensates $\omega_{a}$ found at
the tree level get altered when the high order quantum corrections are
included and we take into account of the set of coupled renormalized
constraint equations in the light-front quantized theory.

It is possible, in the above discussion, to allow for an $\bar x$
dependence in $\omega_{a}$. The first term in the constraint equations
now gets altered and the tree level configurations are now obtained from
$\,[\,V'_a(\omega)-\partial_i\partial_i \omega_a\,]=0$. As before there is
agreement with the Lagrangian formulation. We obtain then the famous
{\it kink} solutions but with an important difference. In the {\it front form}
dynamics the equation for {\it kinks} depends only on the two transverse
directions and not three as in the {\it instant form} case. In two dimensions
theories there are then no kink solutions. This reinforces the affirmation
that the vacuum of the light-front quantized theory is simpler than that
in the {\it instant form}. Moreover, the Hamiltonian maintains locality with
respect
to the transverse directions but it is nonlocal with respect to the
longitudinal direction because of the presence of the constraint equations.
In the presence of fermionic (or other) fields interacting with the scalar
fields the constraint equations would relate the fermionic
condensates with the bosonic ones.

We may also consider the still more general case [10] where $\omega_{a}=
\omega_{a}(\tau,\bar x)$. In order to implement the constraint
$\beta_{a}(\tau,\bar x)\approx 0$ and define the brackets  $\{,\}^{*}$
we need now to invert the matrix

$$C_{ab}({\bar x},{\bar y})\equiv
\{\beta_{a}({\bar x}),p_b({\bar y})\}\,
=\Bigl[\,L[-\,\delta_{ab}\,\partial_i\partial_i
+V''_{ab}(\omega)\,]+V'''_{abc}(\omega)\int dx\,
\varphi_{c}+...\Bigr]\,\delta^{2}({\bar x}-{\bar y}).\eqno(2.52)$$
\vskip 0.2cm
\nl When  $\omega_{a}$,  given by
$\,\omega_{a}(\lambda\omega^{2}-m^{2})=0\,$, are zero, the leading
term, since  $L\to\infty$,  of the matrix  $C$ is
 $\,-L(\partial_{i}\partial_{i}
+m^2)\,\delta_{ab}\delta^{2}({\bar x}-
{\bar y})\,$, while for the case of  $\,(\lambda\omega^{2}-m^{2})=0\,$,
the leading term is
$\,L\,[-\delta_{ab}\partial_{i}\partial_{i}
+2m^2\,P_{ab}\,]\delta^{2}({\bar x}-{\bar y})\,$,
where  $\,P_{ab}=(\omega_{a}\omega_{b})/\omega^{2}\;$  is a projection
operator. In both the cases the inverse of the leading term contains a well
defined Green's function multiplied by an explicit factor of $1/L$.
Consequently, in the continuum, we fall back to the situation similar to that
discussed above in connection with the theory in two dimensions. The final
conclusions then coincide with those obtained in the beginning of
this Section.
It is interesting to note that we may now give a new proof in favour of
the absence of the Goldstone bosons in two dimensions [23]
({\it Coleman's theorem}): Since we have no transverse directions in two
dimensions, the matrix
 $C_{ab}\to 2Lm^{2}P_{ab}\,$. It cannot be inverted and we are unable to
 implement the constraints  $\beta_{a}=0$.

\bigskip
\noindent {\bf 3-  Phase transition in $(\phi^{4})_{2}$ theory:}
\medskip

We will discuss now the  stability of the vacuum in the $\phi^{4}$ theory
when the coupling constant is increased from vanishingly small values to
larger values. The light-front framework seems very appropriate to study
this problem. On renormalizing the theory we have here at our disposal,
in addition to the usual equations like the mass renormalization
condition in the equal-time formulation, also the  renormalized constraint
equations in the theory. For  super-renormalizable
theories in two and three dimensions these eqs. will be shown to contain
all the information needed to study the problem at hand and  they can  also
describe [24] the phase transition. We showed in Sec. 2 that the same
physical description of the spontaneous symmetry breaking is obtained whether
we use the front form or instant form dynamics in spite of the different
mechanisms in the two cases. The same is seen to be true for the problem
we consider below.

We recall that there are rigorous proofs [25] on the triviality of
$\phi^{4}$ theory in the continuum for more than four space time
dimensions and on its interactive nature for dimensions less than four when
the theory is also super-renormalizable. In exactly four dimensions the
situation is still not clarified [25] and it will be very interesting to
study this case on the light-front since it is pertinent to the Higgs sector
of the Standard Model of Weinberg and Salam [26]. However, due to the
complexity of the renormalization problem in four dimensions
we will illustrate our points
considering only the two dimensional theory. From the well established
results on the generalized Ising models, Simon and Griffiths [19]
conjectured some time ago that the two dimensional $\phi^{4}$ theory should
show a {\it second order} phase transition. We do seem to verify this
conjecture by studying the theory quantized on the light-front. Variational
methods like the Hartree approximation or Gaussian effective potential
[27], using {\it front form} but ignoring the constraint [28], or the
one based on a scheme of canonical transformations [29] all lead to a first
order
phase transition contradicting the conjecture. The post-Gaussian
approximation [30] and the non-Gaussian variational method [31] give
a second order transition for a particular value of the coupling constant.
Our result shows it to be of second order for any coupling above a critical
coupling as implied by the mathematical theorem [19]. Our procedure uses
the well established Dyson-Wick expansion [13] of perturbation theory and
may be improved systematically computing still higher order corrections
which is not possible to do in the variational methods. From the
considerations on  the light-front quantized theory we find that we may not
ignore certain contributions in the theory originating from a finite
renormalization corrections. If we drop them our results are in complete
agreement with
those obtained in the variational methods.

\bigskip
\noindent {\bf Renormalization. Phase transition  in two dimensions:}

\medskip

On the light-front we need to renormalize the theory with the Hamiltonian

$$H^{l.f.}=\int d^{2}x \,\Bigl[{1\over2}(m_{0}^2+3\lambda\omega^2)
\varphi^2+\lambda \omega \varphi^3+
{\lambda \over 4}\varphi^4 +{1\over 2}m_{0}^2 \omega^2+{1\over 4}\lambda
\omega^4 \Bigr],\eqno(3.1)$$

\nl in the presence of the constraint equation

$$\,\omega\,(\lambda\,\omega^{2}+m_{0}^{2})\,+\,\lambda\,
lim_{L\to\infty}\,{1\over L}\,\int_{-L/2}^{L/2}
dx\,\,\lbrack 3 \,\omega\,\varphi^{2}+
\varphi^{3}\,\rbrack\,=\,0\,.\eqno(3.2)$$

\nl and the light-front commutator obtained in Sec. 2. It is clear that
it is not convenient to eliminate  $\omega$ using (3.2) since the resulting
Hamiltonian would be quite involved. However, we may renormalize the  theory
based on (3.1) and obtain thereby a renormalized constraint equation.
We have taken here the {\it correct sign}  for the mass term and
   $\,m_{0}\,$ indicates the bare mass which is assumed to be nonvanishing.
We recall that there is no physical consideration in the light-front
framework to normal order the constraint equation. In view of
 $\,k\ne 0\,$ we have $\,\tilde\varphi(k=0)=0\,$ (see also
ref. [32] which implies  $\,\int dx \varphi=0\,$.


We write  $M_{0}^{2}(\omega)= (m_{0}^{2}+3 \lambda \omega^2)$
and choose   ${\cal H}_0={M_0}^2\varphi^2/2$
so that ${\cal H}_{int}= \lambda\omega\varphi^3+\lambda
\varphi^4/4$. The theory being super-renormalizable we need to perform only the
mass renormalization. We could follow as is usually done [33] the old
fashioned noncovariant perturbation theory. However, we could as well use the
covariant propagator (as shown in Appendix A) and use the
Dyson-Wick expansion [13] based on the Wick theorem for exponentials ordered
with respect to light-front time $\tau$

$$T[e^{i\int d^{2}x\, j(x)\varphi(x)}\,]=e^{-{1\over 2}\int\int d^{2}x
 d^{2}y \,j(x)G_{0}(x-y)j(y)}:\,[e^{i\int d^{2}x \,j(x)\varphi(x)}]:\,,$$

\nl where $G_{0}$ is the free propagator of the scalar field.

The  self-energy correction to  {\it one loop order} is

$$\eqalignno{-i\Sigma(p)&=-i\Sigma_1-i\Sigma_2(p) \cr
&=(-i6\lambda){1\over 2} D_{1}({M_0}^2)+(-i6\lambda\omega)^2 {1\over 2}
(-i) D_2(p^2,{M_0}^2)\,,
&(3.3)\cr} $$

\noindent where the divergent contribution  $D_1$ (see  (3.5) below)
refers to the tadpole graph while the one-loop second term comes from the
cubic interaction vertex and gives a finite contribution (Appendix A)
with a sign opposite to that of the first. Also the symmetry
and other  factors from the vertices are  explicitly written.
We argue below that due to the presence of $\omega$ in the second term it
is of the same order in $\lambda$ as the first one and thus cannot be dropped
in the one-loop order we will be considering. This term is also quite
relevant for determining the nature of the phase transition. We remind that
$\langle\varphi(x)\rangle=0$ and the one particle reducible
graphs originating from the cubic term in the interaction are ignored. The
divergences will be handled by the dimensional regularization and we adopt the
minimal subtraction (MS) prescription [34].

The physical mass  $M(\omega)$ is defined [13,34] by

$${M_{0}}^2(\omega)+\Sigma(p)\vert_{p^2=-M^2(\omega)}
= M^2(\omega)\,\eqno(3.4)$$

\noindent where  $p^{\mu}$ is the Euclidean space 4-vector and
$M(\omega)$ determines the pole of the renormalized propagator. Following the
well known procedure of dimensional regularization  we have

$$\eqalignno{D_1(M_0)&= {1\over {(2\pi)}^n}
\int {d^{n}k\over (k^2+{M_0}^2)}\cr &= \mu^{(n-2)}{1\over {4\pi}}
({{M_0}^2\over {4\pi\mu^2}})^{({n\over
2}-1)}\Gamma(1-{n\over2})\cr &\to {\mu^{(n-2)}\over {4\pi}}[{2\over (2-n)}
-\gamma-ln({{M_0^2}\over {4\pi\mu^2}})],\,&(3.5)\cr}$$

\noindent where the limit   $n\to 2$ is understood.
{}From (3.3-5) we obtain

$$ {M_0}^{2}(\omega)=M^{2}(\omega)+{{3\lambda}\over
{4\pi}}\Bigl[\gamma+ln({M^2(\omega)\over {4\pi\mu^2}})
\Bigr]+18\lambda^2\omega^2 D_2(p,M^2)\vert_{p^2=-M^2}
+{3\lambda\over {2\pi}}{1\over {(n-2)}}.\,\eqno(3.6)$$

\noindent Here we have taken into account that in view of the tree level
result
$\,\omega(\lambda\omega^2+{m_0}^2)=\omega[{M_0}^2(\omega)-2\lambda\omega^2]
=0$ the term  $\lambda^2\omega^2 $ ( when $\omega \ne0$) is, in fact, of the
first order in
$\lambda$ and not of the second. We keep the terms only up to first order
in $\lambda$. We remind, however, that
$M_{0}$ depends on $\omega$ and which in its turn is involved in the
constraint equation  (3.2). The expression of
$D_{2}$ is given in the Appendix A; it is finite with the value
 $D_2(p^2,M^2)\vert_{p^2=-M^2}= {\sqrt 3}/(18M^2)$.
To maintain consistency we also replace $M_0$ by $M$
in the terms which are already  multiplied by  $\lambda$.


{}From  (3.6) we obtain the {\it mass renormalization condition }

$$M^2-m^2=3\lambda\omega^2+{{3\lambda}\over {4\pi}}
ln({m^2\over {M^2}})-\lambda^2
\omega^2 {\sqrt{3}\over {M^2}}\,\eqno(3.7)$$

\noindent where for convenience of writing we have set $M(\omega)\equiv M$ and
$M(\omega=0)\equiv m$ indicating the physical masses in the
asymmetric and symmetric phases respectively.
The eq. (3.7) expresses the invariance of the bare mass ${m_0}^2$.
For $\omega=0$ or $\lambda=0$ it implies $M^2=m^2$ ({\it symmetric phase}).

Consider next the constraint equation (3.2). To the lowest order
we find [13,20]

$$\eqalignno{\quad{3\lambda\omega\langle\varphi(0)^2\rangle
&\simeq 3\lambda \omega.iG_{0}(x,x)}= 3\lambda\omega. D_1(M),\cr
{\lambda\langle\varphi(0)^3\rangle &\simeq \lambda
(-i{\lambda}\omega).6.\int dx
\langle T(\varphi(0)^{3}\varphi(x)^{3})\rangle_{c}^{0},}\cr
&= -6\lambda^2\omega D_3(M)=-6\lambda^2\omega {b\over{(4\pi)^2 M^2}}
\,,&(3.8)\cr}$$

\noindent where {\it c} indicates {\it connected} diagram and $D_3$
(Appendix A) is a finite integral with $b\simeq 7/3$.
Taking the vacuum expectation value of
the constraint equation (3.2) and on making use of (3.5-8)
we find that the divergent term cancels and we obtain the
{\it renormalized constraint equation}

$$\beta(\omega)\equiv\omega\Bigl[M^2-2\lambda\omega^2 +
\lambda^2\omega^2 {{\sqrt 3}\over M^2}
-{{6\lambda^2}\over (4\pi)^2}{b\over M^2}\Bigr]=0.\,\eqno(3.9)$$


We will verify below that $\beta$  coincides with
the total derivative with respect to $\omega$,
in the equal-time formulation, of the  (finite) difference $F(\omega)$ (see
below)
of the  renormalized vacuum energy densities  in the {\it asymmetric}
($\omega\ne0$) and
{\it symmetric} ($\omega=0$) phases in the theory. The last term in $\beta$
corresponds to a correction $\simeq \lambda(\lambda\omega^2)$ in this energy
difference and thus may not be ignored just like in the case of the
self-energy discussed above.
In the equal-time case   (3.9) would be
required to be {\it added} to the theory
upon physical considerations. It will  ensure that
the sum of the tadpole diagrams, to the approximation concerned,
for the
transition $\varphi\to vacuum$ vanishes. The physical outcome would then be
the same in the two forms of treating the theory here  discussed.
The variational methods write only the first two ($\approx$ tree level)
terms in the  expression for $\beta$  and thus ignore
the terms coming from the finite corrections.
A similar remark can be made about the last term in (3.7).
Both of the eqs. (3.7) and (3.9) and
the difference of energy densities above are also found
to be independent of the arbitrary mass scale introduced in
the dimensional regularization and  contain only the finite
physical parameters of the theory.  The finite
renormalization corrections alter the critical
coupling and the nature of the phase transition compared to what found
in the case of the variational methods.

Consider first the {\it symmetric phase} with $\omega\approx 0$,
which is allowed by (3.9), and for which
$M^2\approx m^2 $ as follows from (3.7). The latter
also allows us to compute
$\partial M^2/\partial{\omega}=
2\lambda\omega(3-{\sqrt 3}\lambda/M^2)/[1+3\lambda/(4\pi M^2)-{\sqrt 3}
\lambda^2\omega^2/M^4]$ which is needed to find  $\beta'\equiv
d\beta/d\omega$. Its  sign will determine the nature of the stability
of the vacuum corresponding to a particular value of $\omega$
obtained from (3.9). In the symmetric phase we obtain
$\beta'(\omega=0)=M^2[1-0.0886(\lambda/M^2)^2]$.
It changes the sign from a positive value for vanishingly weak couplings to
a negative value when the coupling increases.  In other words
the system starts out in a stable symmetric phase for very small coupling but
goes over into an unstable symmetric phase for values above the small coupling
$g_{s}\equiv\lambda_{s}/(2\pi m^2)\simeq 0.5346$.


Consider next the case of {\it the spontaneously broken symmetry
phase} ($\omega\ne 0$). It follows from (3.9)
that the nonzero values of $\omega$ are found from

$$M^2-2\lambda\omega^2
+{{\sqrt 3}\lambda\over 2}=0,\,\qquad\qquad (\omega\ne0),\eqno(3.10)$$

\noindent where we made use of $\, 2\lambda \omega^2\simeq M^2\,$
in the zero order approximation when $\omega\ne 0$.
The mass renormalization condition now reads as

$$M^2-m^2=3\lambda\omega^2+{{3\lambda}\over {4\pi}}\,
ln({m^2\over {M^2}})-\lambda {\sqrt{3}\over2},\eqno(3.11)$$

\noindent On eliminating $\omega$ from (3.10),(3.11)
we obtain the {\it modified duality relation}

$${1\over {2}}M^2+m^2+{{3\lambda}\over {4\pi}}\,
ln({{m^2}\over {M^2}})+{{\sqrt 3}\over 4}\lambda=0.\,\eqno(3.12)$$

\noindent which can also be rewritten as
$[\lambda\omega^2+m^2+(3\lambda/({4\pi}))
ln(m^2/M^2)]=0$ and it shows that  the  real solutions exist only for
$M^2 > m^2$. The finite corrections found here are again
not considered in the references cited in Sec. 1, for
example, they  assume (or find)  the tree level expression
$\,M^{2}-2\lambda\omega^{2}=0$.
In terms of  the dimensionless coupling constants
$g=\lambda/(2\pi m^2)\ge 0$ and $G=\lambda/(2\pi M^2)\ge 0$
we  have  $G<g$.
The new self-duality eq. (3.12) differs from the old one [27,28] and
shifts  the critical coupling to a higher value.
We find that: {\it i)} for $g < g_c=6.1897$ ({\it critical coupling})
there is no real solution for $G$, {\it ii)} for a fixed $g>g_{c}$ we have two
solutions for $G$ one with the point lying on the upper branch
($G>1/3$) and the other with that on the lower branch ($
G<1/3$),  of the curve describing $G$ as a function of $g$ and
 which starts at the point $(g=g_{c}=6.1897,G=1/3)$,
{\it iii)} the lower branch with $G<1/3$,
approaches to a vanishing value for $G$ as $g\to\infty$, in contrast to the
upper one for which $1/3<G<g$ and  $G$ continues to increase.
{}From (3.11) and   $\beta=\omega[M^2-2\lambda\omega^2
+{{\sqrt 3}\lambda/2}]\,$ we  determine
$\beta'\approx (1+0.9405 G)$ which is always  positive and thus
indicates a minimum of the difference of the vacuum energy densities
for  the nonzero values of $\omega$.

The energetically favored broken symmetry phases
become available only after the
coupling grows to the critical coupling $g_c=6.18969$
and beyond this the asymmetric phases would be preferred against
the unstable symmetric phase in which the system finds itself when
$g>g_{s}=0.5346$. The phase transition is thus of the {\it second order}
confirming the conjecture of Simon-Griffiths. If we ignore the additional
finite renormalization corrections the well known results following
from the variational methods are  reproduced exactly in our calculation,
e.g., the symmetric phase always remains stable but for $g>1.4397$ the
energetically favored asymmetric phases also do appear, indicating a first
order transition.

\bigskip

\noindent {\bf Vacuum energy density:}
\medskip
The expression for the vacuum energy density in the conventional equal-time
formulation is given by

$${\cal E(\omega)}=I_{1}(M_0)+{1\over 2}{m_0}^2\omega^{2}+\,{\lambda\over 4}
\omega^{4}+{{\lambda}\over 4}.3.{D_1(M_0)}^2+{(-i6\lambda\omega)^2}.{1\over
{2!}}.
 {1\over 6}.{D_3}(M_0),\eqno(3.13)$$

\noindent where the symmetry and other factors are explicitly written.
The first term  is the energy density with respect to the free propagator
with mass ${M_{0}}^{2}$ and is given by [27]

$$\eqalignno{I_1(M_0)&={1\over (2\pi)^{(n-1)}}\int d^{(n-1)}k\,\,
{1\over 2}\,\,\sqrt{\vec k^{2}+{M_0}^2}\cr
&={{M_0}^{n}\over {(4\pi)^{n\over2}}}\,\,{1\over n}\,\,\Gamma(1-{n\over 2})
\cr &\to {\mu^{(n-2)}}\,{{M_0^2}\over {4\pi}}\,{1\over 2}\,
\Bigl[\,{2\over {(2-n)}}+
1-\gamma-ln({{M_0}^2\over{4\pi\mu^2}})\,\Bigr]\,&(3.14)\cr}$$

\nl The ${D_{1}}^{2}$ term represents the two-loop correction of the order
$\lambda $ in the coupling constant and so does  the last one in view of the
discussion above except for that it gives a finite contribution and
carries an opposite sign. We remark that the  last term
is non-vanishing even in the  light-front framework. Here we find
in the integrand  $\theta(k)\theta(k')\theta(k'')\delta(k+k'+k'')$
multiplied by another distribution. The product distribution, however,
may not be considered vanishing. The last term of (3.9) corresponds to the
derivative with respect to $\omega$ of the last term in (3.13). In the
variational methods this term is found ignored.

A finite expression
for the difference of the vacuum energy densities
in the broken  and the symmetric phases,  which is also
independent of the arbitrary mass $\mu $ introduced in
the dimensional regularization, is obtained to be [24]

$$\eqalignno{F(\omega)&={\cal E}(\omega)-{\cal E}(\omega=0)\cr
&={(M^2-m^2)\over {8\pi}}+{1\over {8\pi}}(m^2+3\lambda\omega^2)
\,ln({m^2\over M^2})+{3\lambda\over 4}\Bigl[{1\over {4\pi}}ln({m^2\over M^2})
\Bigr]^2\,\cr
&\qquad\quad+{1\over 2}m^2 \omega^2+{\lambda\over 4}\omega^4+{1\over {2!}}.
(-i6\lambda\omega)^2.{1\over 6}.D_3(M)\quad&(3.15)\cr} $$

\noindent We verify that $(dF/d\omega)=\beta$
and $d^2F/d^2\omega=\beta'$ to the one-loop order. Except for the last term
it coincides with the results in the earlier works.
{}From numerical computation we verify that at the minima
corresponding to the nonvanishing value of $\omega$
the value of F is negative
and that for a fixed $g$ it is more negative for the point on the
lower branch ($G<1/3$) than for that on the upper branch
($G>1/3$).  To illustrate  we find:
for $g=6.366$ and $ G=0.263$ we get $\vert\omega\vert
=0.736,\, F=-0.097\lambda $
while for the same $g$ but $G=0.431$ we find $\vert\omega\vert=0.617,
\, F=-0.082\lambda $. For $g=11.141$ and $ G=0.129$ we get $\vert
\omega\vert=1.050,\,
F=-0.174\lambda $
while for the same $g$ but $G=1.331$ we find $\vert\omega\vert=0.493,\,
F=-0.111\lambda $.  The symbolic manipulation is convenient to handle
(3.7) and (3.9).

\bigskip

\nl {\bf 4- Conclusion:}
\medskip
The  present work and the earlier one
on the mechanism of spontaneous continuous symmetry breaking [9,10] add
to the previous experience [1-8] that the {\it front form} dynamics is a useful
complementary method and needs to be studied systematically in the context of
QCD and other problems. The physical results following from
one or the other form of the theory
should come out to be the same though the mechanisms to arrive at them
may be  different. In the equal-time case
we  introduce external considerations in order
to constrain the theory
while  the analogous conditions in the light-front formulation
seem to be contained in it through the  self-consistency equations.
In the discussion of the phase transition in  $\phi^{4}$
theory the finite renormalization
corrections should also be taken in to account. In any case both
the light-front and  the equal-time (space like) hyperplanes are equally
valid for formulating the field theory dynamics.

\bigskip
\bigskip
\noindent{\bf Acknowledgements:}
\medskip

The author acknowledges
the hospitality offered to him by the {\sl Physics Department,
Ohio State University}, Columbus, Ohio,
and by the {\sl INFN and  Dipartimento di Fisica, Universit\`a di
Padova}, Padova, where  the work presented here was completed. A research
grant from INFN-Padova is gratefully acknowledged.
Acknowledgements with thanks are due to Avaroth Harindranath and Robert Perry
for many constructive suggestions and  clarifications
during the progress of the work. Acknowledgement with thanks
are also due to Ken Wilson for  asking
useful questions and several constructive suggestions,
to Stan Brodsky for
encouragement,  to Mario Tonin, Stuart Raby, G. Costa,
A. Bassetto, P. Marchetti, and G. Degrassi  for several discussions.

\vfill \eject

\centerline {\bf Appendix A}
\medskip
\noindent {\bf Light-front propagator:}
\vskip 0.3cm
The propagator for the free (e.g., in the interaction representation)
massive scalar field is defined by
$iG_{0}(x;x')\equiv\langle 0\vert T\{\varphi(x,\tau)
\varphi(x',\tau')\}\vert 0\rangle$, where T indicates
the ordering in light-front time $\tau$. On using the  momentum space
expansion and
commutation relations of the operators we arrive at

$$iG_0(x,0)={1\over {2\pi}}\int\,
{dk\over {2k}}\theta(k)\Bigl[\theta(\tau)\,
e^{-i(kx+\epsilon_k\tau)}+\theta(-\tau)\,e^{i(kx+\epsilon_k\tau)}\Bigr],
\eqno (A.1)$$

\noindent where $\,2k\epsilon_{k}=m^{2}$. Consider next the usual Feynman
propagator, where we change
the variables from $\,k^0,k^1$ to $k^+=(k^0+k^1)/{\sqrt 2},\,k^-=
(k^0-k^1)/{\sqrt 2}$ with $-\infty<k^+,k^-<\infty$,

$$\int\int
{{d{k^{+}}d{k^{-}}}\over {(2\pi)^2}}{i\over (2k^{+}k^{-}-m^2+i\epsilon)}
\,[\theta(k^{+})+\theta(-k^{+})]\,\, e^{-i(k^{+}x+k^{-}\tau)},\eqno(A.2) $$

\noindent and where we have inserted the identity
$1=[\theta(k^{+})+\theta(-k^{+})]$  valid in the sense of
distribution theory. We rewrite this expression as

$$\int\int
{{d{k^{+}}d{k^{-}}}\over {(2\pi)^2}}\,i\,\,\Bigl[\,
{{ e^{-i(k^{+}x+k^{-}\tau)} \,\theta(k^+)}
\over {2k^{+}\, (k^{-}-\epsilon(k^{+})+i\epsilon)}}+\,
{{ e^{-i(k^{+}x+k^{-}\tau)}\, \theta(-k^+)}
\over {2k^{+}\, (k^{-}-\epsilon(k^{+})-i\epsilon)}}\,\Bigr]$$

\noindent where $\eta(k^{+})=m^2/(2k^{+})$.
Next we make the change $k^+\to -k^+, k^- \to -k^-$ in the
second term to recast it as

$$\int\int
{{d{k^{+}}d{k^{-}}}\over {(2\pi)^2}}i\theta(k^+)}\Bigl[
{{ e^{-i(k^{+}x+k^{-}\tau)}
\over {2k^{+}\, (k^{-}-\eta(k^{+})+i\epsilon)}}+
{{ e^{+i(k^{+}x+k^{-}\tau)}
\over {2k^{+}\, (k^{-}-\eta(k^{+})+i\epsilon)}}\Bigr]$$

\noindent If we make the {\it rule}
that the $k^{-}$ integration has to be
performed first we obtain, on using the well known
integral representations of $\theta(\tau)$, the
light-front propagator $(A.1)$. Inversely we could introduce
these representations directly in $(A.1)$ and
arrive at $(A.2)$.
In the gauge theories with the infrared singularities also present
we need to find  an adequate procedure to regulate (subtractions)
the integrals in momentum space before so as to
ensure that the $k^-$ integration may be performed first. We recall
that also in the equal-time formulation similar arguments requiring that
the $k^{0}$ integration be done first are made [35].

\bigskip

\noindent {\bf Integrals $D_2,\, D_3$:}
\medskip
The finite integrals appearing in the text are well known and
easily computed after transforming them to the Euclidean space integrals
by Wick rotation as usual

$$D_{2}(p^2,{M_0}^2)&=\int {d^2k\over {(2\pi)^2}} {1\over
{{(k^2+M_0^2)}{[(p-k)^2+{M_0}^2]}}}\eqno(A.3) $$

$$\eqalignno{D_3(M)&=\int\int {d^2k\over {(2\pi)^2}}
{d^2q\over {(2\pi)^2}}, {1\over
{(k^2+M^2)(q^2+M^2)[(q+k)^2+M^2]}},\cr &={1\over {(4\pi)^2 M^2}}
\int_{0}^{1}\int_{0}^{1}dx dy {1\over
{[1-y+xy(1-x)]}}\equiv {b\over(4\pi)^2 M^2}\,, &(A.4)\cr}$$

\noindent We find $D_2(p^2=-{M_0}^2,{M_0}^2)={\sqrt 3} /(18{M_0}^2)$.
We could alternatively perform the corresponding computation using the
propagators in (A.1) and follow the old fashioned perturbation theory [36]
obtaining the same results.

\bigskip
\bigskip
\nl {\bf References:}
\medskip

\item{[1.]}P.A.M. Dirac, Rev. Mod. Phys. {\bf 21} (1949) 392.
\item{[2.]}S. Weinberg, Phys. Rev. {\bf 150} (1966) 1313.
\item{[3.]}J.B. Kogut and D.E. Soper, Phys. Rev. {\bf D 1} (1970) 2901.
\item{[4.]}S. Fubini, G. Furlan, and C. Rossetti, Nuovo Cimento {\bf 40}
(1965) 1171; S. Fubini, Nuovo Cimento {\bf 34A} (1966) 475;
R.F. Dashen, M. Gell-Mann: Proc. of the 3rd Coral Gable Conference on
Symmetry Principles at High Energy, W.H. Freeman and Co., San Francisco, 1966.
\item{[5.]}K. Bardacki and G. Segre, Phys. Rev. {\bf 153} (1967) 1263;
J. Jersak and J. Stern, Nuovo Comento  {\bf 59} (1969) 315;
H. Leutwyler, Springer Tracts Mod. Phys. {\bf 50} (1969) 29;
R. Jackiw, Springer Tracts Mod. Phys. {\bf 62} (1972) 1;
F. Rohrlich, Acta Phys. Austr. {\bf 32} (1970) 87.
\item{[6.]}D. Gross, J. Harvey, E. Martinec, and R. Rohm, Phys. Rev. Lett.
{\bf 54} (1985) 503; Nucl. Phys. {\bf B256} (1985) 253.
\item{[7.]}K.G. Wilson, Nucl. Phys. B (proc. Suppl.) {\bf 17} (1990);
R.J. Perry, A. Harindranath, and K.G. Wilson,
Phys. Rev. Lett. {\bf 65} (1990)  2959; S. Glazek, A.H. Harindranath,
S. Pinsky, J. Shigemitsu, and K. Wilson, Phys. Rev. {\bf D47} (1993) 1599.

\item{[8.]}S.J. Brodsky and H.C. Pauli, {\it Schladming Lectures}, {\sl SLAC}
 preprint {\sl SLAC}-PUB-5558/91; K. Hornbostel, preprint, Cornell
University, CLNS 91/078 (1991);
H.C. Pauli and S.J. Brodsky, Phys. Rev. {\bf D 32} (1985) 1993
and 2001; Phys. Rev. {\bf D 32} (1985) 2001;
S.J. Brodsky and G.P. Lepage, in {\it Perturbative Quantum
Chromodynamics}, edited by A.H. Mueller, World Scientific, Singapore, 1989.

\item{[9.]} P.P. Srivastava, {\it Spontaneous symmetry breaking mechanism
in light-front quantized field theory}- {\sl Discretized formulation},
 Ohio State University preprint  92-0173, {\sl SLAC} PPF-9222, April 1992- {\sl
Continuum limit of the discretized formulation is obtained. The non-local
nature of light-front Hamiltonian is pointed out}. Contributed
to {\it XXVI Int. Conf. on High Energy Physics, Dallas, Texas},
August 92, (AIP Conference Proceedings 272, Ed. J.R. Sanford ).

\item{[10.]} P.P. Srivastava, {\it On spontaneous symmetry breaking
mechanism in light-front quantized
field theory}, Ohio State University preprint 91-0481, {\sl SLAC} database no.
PPF-9148, November 1991 e CBPF-NF-004/92- {\sl continuum formulation
in two dimensions.};
\item{}{\it Higgs mechanism ({\sl tree level}) in light-front
quantized field theory},  Ohio State University preprint 92-0012,
{\sl SLAC} PPF-9202, December 91, CBPF-NF-010/92- {\sl
continuum formulation in
$3+1$ dimensions}. Contributed
to {\it XXVI Int. Conf. on High Energy Physics, Dallas, Texas},
August 92, (AIP Conference Proceedings 272, Ed. J.R. Sanford );
\item{}{\it Constraints in light-front quantized field theory},
 Ohio State University preprint 92-0175,
{\sl SLAC} PPF-9222, December 91.  Contributed
to {\it XXVI Int. Conf. on High Energy Physics, Dallas, Texas},
August 92, (AIP Conference Proceedings 272, Ed. J.R. Sanford );
\item{}{\it Constraints and Hamiltonian
in light-front quantized theory},
Padova University preprint, DFPF/92/TH/58, December 92, Nuovo Cimento A.

\item{[11.]}P.A.M. Dirac, {\it Lectures
in Quantum Mechanics}, Benjamin, New York, 1964; E.C.G. Sudarshan and
N. Mukunda, {\it Classical Dynamics: a modern perspective}, Wiley, N.Y.,
1974; A. Hanson, T. Regge and C. Teitelboim, {\it Constrained
Hamiltonian Systems}, Acc, Naz. dei Lincei, Roma, 1976.

\item{[12.]}H. Lehmann, Nuovo Cimento {\bf 11} (1954) 342. See also
[13].
\item{[13.]} S.S. Schweber, {\it An Introduction to
Relativistic Quantum Field Theory}, Harper and Row, Inc., New York, 1962;
J.D. Bjorken and S.D. Drell, {\it Relativistic Quantum
Fields}, McGraw-Hill, 1965; D. Luri\'e, {\it Particles and Fields},
Interscience Pub., New York, 1968.

\item{[14.]}T. Maskawa and K. Yamawaki, Prog. Theor. Phys. {\bf 56} (1976)
270.
\item{[15.]}N. Nakanishi and K. Yamawaki, Nucl. Phys. {\bf B122} (1977) 15.
\item{[16.]}R.S. Wittman, in Nuclear and Particle Physics
on the Light-cone, eds.
M.B. Johnson and L.S. Kisslinger, World Scientific, Singapore, 1989.
\item{[17.]}Th. Heinzl, St. Krusche, and E. Werner,
Phys. Lett. {\bf B 256} (1991) 55; Th. Heinzl, St. Krusche, and
E. Werner, Regesburg preprint, TPR 91-23; Phys. Lett. {\bf B 272} (1991) 54.
\item{[18.]}G. McCartor and D.G. Robertson, SMU preprint SMUHEP/91-04.

\item{[19.]}B. Simon and R.B. Griffiths, Commun. Math. Phys. {\bf
33} (1973) 145;  B. Simon, {\it The ${P(\Phi)_{2}}$ Euclidean (Quantum)
Field Theory}, Priceton University Press, 1974;
J. Glimm and A. Jaffe, {\it Quantum Physics: A functional integral point
of view}, Springer, New York, 1981 and references therein.

\item{[20.]} See for example,  G. Parisi, {\it Statistical Field Theory},
Addison-Wesley, 1988; G. Barton, {\it Introduction to Advanced
Field Theory}, p. 135, Interscience Pubs. (John Wiley), 1963.

\item{[21.]} S. Chang, R. Root, and T. Yan, Phys. Rev. {\bf D7} (1973)
1133, 1147; S. Chang and S. Ma, Phys. Rev. {\bf 180} (1969) 1506;
T. Yan, Phys. Rev. {\bf D7} (1973) 1780.

\item{[22.]}S. Weinberg, Phys. Rev. Lett.  {\bf 29} (1972) 1698.

\item{[23.]}S. Coleman, Commun. Math. Phys. {\bf 31} (1973) 259.

\item{[24.]}P.P. Srivastava, {\it Light-front field theory and nature of phase
transition in
$({\phi^{4}})_{2}$ theory}, Padova university preprint DFPF/93/TH/18, March
1993.

\item{[25.]} D. Brydges, J. Fr\"olich and A.D. Sokal, Commun. Math. Phys.
{\bf 91} (1983) 141;
M. Aizenman, Phys.Rev. Lett. {\bf 97} (1981) 1;
Commun. Math. Phys. {\bf 86} 1; J. Fr\"ohlich, Nucl. Phys. {\bf B200 [FS4]}
(1982) 281;
D.J.F. Callaway, Phys. Rep. {\bf 167} (1988) 241.

\item{[26.]}S. Weinberg, Phys. Rev. Lett. {\bf 19} (1967) 1264;
A. Salam, Proc. 8th Nobel Symposium, Aspen\"asgarden, 1968, ed.
N. Svartholm (Almqvist and Wiksell, Stockholm, 1968), p.367.

\item{[27.]}The topic has a long history.  See for example,

\item{}P.M. Stevenson, Phys. Rev. {\bf D 32} (1985) 1389,
\item{}M. Consoli and A. Ciancitto, Nucl. Phys. {\bf B254} (1985) 653.
\item{} and the references cited therein.
\item{}S.J. Chang, Phys. Rev. {\bf D 13} (1976) 2778,
\item{[28.]}A. Harindranath and J.P. Vary, Phys. Rev. {\bf D 37}
(1985) 1389.
\item{[29.]}G.V. Efimov, Int. Jl. Mod. Phys. {\bf A 4} (1989) 4977.
\item{[30.]}M. Funke, U. Kaulfuss, and H.Kummel,
Phys. Rev. {\bf D 35} (1987) 621.

\item{[31.]}L. Polley and U. Ritschel, Phys. Lett. {\bf B 221}(1989) 44.
\item{[32.]}S. Schlieder and E. Seiler, Commun. Math. Phys. {\bf 25} (1972) 62.

\item{[33.]}A. Harindranath and R.J. Perry, Phys. Rev. {\bf D 43} (1991)
492.

\item{[34.]}J. Collins, {\it Renormalization}, Cambridge University Press,
1984.

\item{[35.]}J.M. Jauch and F. Rohrlich,
{\it Theory of Photons and Electrons}, Addison-Wesley Pub. Co.,
Massachusetts, 1954.

\item{[36.]}W. Heitler, {\it Quantum Theory of Radiation}, Oxford University
Press, fourth edtion.

\bye